\documentclass[pra,twocolumn,superscriptaddress,showpacs,nofootinbibfloatfix,amsmath,amsfonts,amssymb]{revtex4-1}%
\usepackage{amsmath,amsfonts,amssymb,color}
\usepackage{amsthm}
\usepackage{leftidx}
\usepackage{graphicx}
\usepackage{xcolor}
\usepackage{dcolumn}
\usepackage{bm}
\usepackage{epstopdf}
\usepackage{epsfig}
\usepackage{environ}
\usepackage{pdfcomment}

\usepackage{float}
\usepackage[T1]{fontenc}
\usepackage[latin9]{inputenc}
\usepackage{setspace}
\usepackage{esint}

\begin{document}
	
\preprint{APS/123-QED}

\title{Entanglement spectrum and entropy in Floquet topological matter}

\author{Longwen Zhou}
\email{zhoulw13@u.nus.edu}
\affiliation{%
	College of Physics and Optoelectronic Engineering, Ocean University of China, Qingdao, China 266100
}

\date{\today}

\begin{abstract}
Entanglement is one of the most fundamental features of quantum systems.
In this work, we obtain the entanglement spectrum and entropy of
Floquet noninteracting fermionic lattice models and build their connections
with Floquet topological phases. Topological winding and Chern numbers
are introduced to characterize the entanglement spectrum and eigenmodes. 
Correspondences between the spectrum
and topology of entanglement Hamiltonians under periodic boundary
conditions and topological edge states under open boundary conditions
are further established. The theory is applied to Floquet topological insulators in different
symmetry classes and spatial dimensions. Our work thus provides a useful framework
for the study of rich entanglement patterns in Floquet topological matter.
\end{abstract}

\pacs{}
\keywords{}
\maketitle

\section{Introduction\label{sec:Int}}

Floquet topological phases are intrinsically nonequilibrium states
in periodically driven systems \cite{FloRev1,FloRev2,FloRev3,FloRev4,FloRev5}.
They are characterized by large topological invariants \cite{FloBigTN1,FloBigTN2,FloBigTN3,FloBigTN4,FloBigTN5,FloBigTN6,FloBigTN7,FloBigTN8,FloBigTN9,FloBigTN10},
unique symmetry classifications \cite{FloClass1,FloClass2,FloClass3,FloClass4}
and anomalous edge modes with no static analogies \cite{FloAnoES1,FloAnoES2,FloAnoES3}.
The realization of Floquet topological matter in various experimental
settings \cite{FloExp1,FloExp2,FloExp3,FloExp4,FloExp5,FloExp6,FloExp7,FloExp8}
further promotes the development of new methods in quantum engineering
\cite{FloRev2}, ultrafast electronics \cite{FloRev3} and topological
quantum computing \cite{FloQC1,FloQC2,FloQC3}.

Entanglement is one of the most profound concepts in quantum physics
\cite{EntangleRev1,EntangleRev2,EntangleRev3,EntangleRev4,EntangleRev5,EntangleRev6,EntangleRev7}.
It characterizes the non-classical correlation among different parts
of a composite quantum system. Information theoretical measures, such
as the entanglement spectrum (ES) \cite{ES0} and entanglement entropy
(EE) \cite{EE0}, have been further shown to be able to provide important
insights for the understanding of topological phases in condensed matter
systems \cite{ESEETP1,ESEETP2,ESEETP3,ESEETP4,ESEETP42,ESEETP5,ESEETP52,ESEETP6,ESEETP7,ESEETP8,ESEETP9,ESEETP10,ESEETP11,ESEETP12,ESEETP13}.
For example, in two-dimension, topology-induced subleading corrections
were found in the EE \cite{ESEETP3,ESEETP4,ESEETP42}. Moreover, the
ES of reduced density matrix has been shown to contain information
about the bulk-edge correspondence in topological insulators and superconductors
\cite{ESEETP5,ESEETP52,ESEETP6,ESEETP7,ESEETP8,ESEETP9,ESEETP10,ESEETP11,ESEETP12}.
In Floquet systems, the ES and EE have also been considered in periodically
driven Kitaev chain and graphene lattice, and unique bulk-edge correspondences
in the EE of Floquet ground states were identified \cite{YatesPRB2016,YatesPRB2017,YatesPRL2018,JafariPRA2021}.
However, a systematic approach to reveal the ES and EE for general
states of Floquet systems in different physical dimensions still awaits
to be developed. Furthermore, connections between the topological
properties of Floquet entanglement Hamiltonians and Floquet topological edge states
for models belonging to different symmetry classes and with different numbers
of quasienergy bands have yet to be established.

\begin{figure}
	\begin{centering}
		\includegraphics[scale=0.36]{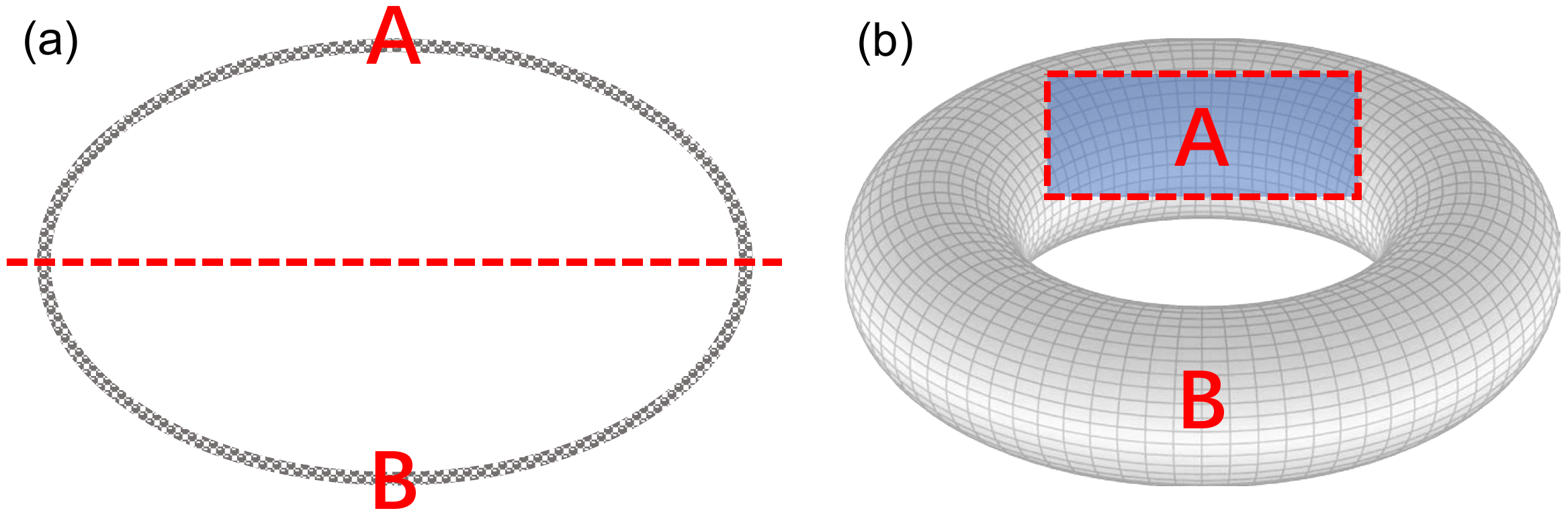}
		\par\end{centering}
	\caption{Partitions for the bipartite ES and EE of 1D chain (a) and 2D
		lattice (b) under periodic boundary conditions. Red dashed lines denote the partition boundaries
		(entanglement cuts) between the subsystems A and B. \label{fig:Sketch}}
\end{figure}

In this work, we introduce a framework to characterize the ES and
EE of Floquet systems consisting of noninteracting fermions in discrete lattices. 
The recipe of our theory is developed in Sec.~\ref{sec:The}, where we build
the connection between the ES and EE of a Floquet system and the spectrum
of its correlation matrix. 
The entanglement is introduced by first decomposing the whole system S into two parts A and B in coordinate space, and then tracing out all the degrees of freedom belonging to the subsystem B from the many-particle density matrix $\hat{\rho}$ of the considered Floquet state in S. The reduced density matrix $\hat{\rho}_{\rm A}={\rm Tr}_{\rm B}\hat{\rho}$ then carries the information about the bipartite entanglement between subsystems A and B. Illustrations of the considered partitions for one- and two-dimensional systems are given in Fig.~\ref{fig:Sketch}. We will refer to the entanglement Hamiltonian $\hat{H}_{\rm A}$ of $\hat{\rho}_{\rm A}$, defined by $\hat{\rho}_{\rm A}=e^{-\hat{H}_{\rm A}}/{\rm Tr}(e^{-\hat{H}_{\rm A}})$ as the Floquet entanglement Hamiltonian, as it is originated from the many-particle Floquet state.
Real-space winding and Chern number invariants
are further introduced to depict the topological nature of Floquet
entanglement Hamiltonians in one and two spatial dimensions. In Sec.~\ref{sec:Res},
we apply our theory to typical one-dimensional (1D)
Floquet topological insulator models in different symmetry classes
and two-dimensional (2D) Floquet Chern insulator models with different
numbers of quasienergy bands. For each case, we establish concrete
relations between the topological properties of the ES under the periodic
boundary condition (PBC) and the topological edge states of the Floquet
operator under the open boundary condition (OBC). These entanglement
bulk-edge correspondences are generic and applicable to other systems
within the same symmetry classes as those considered in this work.
We summarize our results, compare them with related studies and discuss 
potential future directions in Sec.~\ref{sec:Sum}.
More details about the properties of EE around Floquet
topological phase transitions are provided in the Appendix \ref{app:A}.

\section{Theory\label{sec:The}}
In this section, we introduce the definitions of ES and EE for Floquet
states in periodically driven noninteracting fermionic systems. We
further establish their connection with the single-particle correlation
matrix of the system. The latter  provides us with an efficient means to explore
the ES and EE theoretically and numerically. Finally, we introduce topological
invariants for the reduced density matrix, which will be used to characterize
Floquet topological phases at the level of entanglement Hamiltonian
in one and two spatial dimensions. 

In the second quantization formalism, the time-dependent Hamiltonian
for a Floquet system of free fermions reads
\begin{equation}
	\hat{H}(t)=\sum_{m,n}\hat{c}_{m}^{\dagger}H_{mn}(t)\hat{c}_{n},\label{eq:H}
\end{equation}
where $\hat{H}(t)=\hat{H}(t+T)$ and $H_{mn}(t)=H_{mn}(t+T)$ are
periodic in time $t$ with the driving period $T$ and driving frequency
$\omega=2\pi/T$. $\hat{c}_{n}$ ($\hat{c}_{n}^{\dagger}$) is the
annihilation (creation) operator of a fermion in the basis $\{|n\rangle\}$
so that $\hat{c}_{n}|n\rangle=|\emptyset\rangle$ and $\hat{c}_{n}^{\dagger}|\emptyset\rangle=|n\rangle$,
where $|\emptyset\rangle$ denotes the vacuum state. The fermionic creation
and annihilation operators satisfy the anticommutation relations
\begin{equation}
	\left\{\hat{c}_{m},\hat{c}_{n}\right\}=\left\{\hat{c}_{m}^{\dagger},\hat{c}_{n}^{\dagger}\right\}=0,\qquad\left\{\hat{c}_{m},\hat{c}_{n}^{\dagger}\right\}=\delta_{mn}.\label{eq:ACR1}
\end{equation}

The Floquet operator of the system generates its dynamics over a
complete driving period. Consider the evolution from $t=0$ to $T$, 
it can be defined as
\begin{equation}
	\hat{U}=\hat{\mathsf{T}}e^{-\frac{i}{\hbar}\int_{0}^{T}\hat{H}(t)dt},\label{eq:UT}
\end{equation}
where $\hat{\mathsf{T}}$ is the time-ordering operator. From now on we set $\hbar=1$.
Since $\hat{H}(t)$
is Hermitian, $\hat{U}$ is unitary and it possesses a single-particle
basis $\{|\psi_{j}\rangle\}$ called Floquet eigenstates. In this
basis, we can express $\hat{U}$ as
\begin{equation}
	\hat{U}=e^{-i\sum_{j}E_{j}\hat{\psi}_{j}^{\dagger}\hat{\psi}_{j}},\label{eq:U}
\end{equation}
where $E_{j}\in[-\pi,\pi)$ is the quasienergy. The operator $\hat{\psi}_{j}^{\dagger}$
creates a particle in the Floquet state $|\psi_{j}\rangle$ from vacuum,
i.e., $\hat{\psi}_{j}^{\dagger}|\emptyset\rangle=|\psi_{j}\rangle$. $\{|n\rangle\}$
and $\{|\psi_{j}\rangle\}$ are both orthonormal and complete basis
satisfying
\begin{equation}
	\langle m|n\rangle=\delta_{mn},\qquad\sum_{n}|n\rangle\langle n|=\hat{I},\label{eq:Basis1}
\end{equation}
\begin{equation}
	\langle\psi_{i}|\psi_{j}\rangle=\delta_{ij},\qquad\sum_{j}|\psi_{j}\rangle\langle\psi_{j}|=\hat{I},\label{eq:Basis2}
\end{equation}
with $\hat{I}$ being the identity operator. We can thus relate $\hat{\psi}_{j}^{\dagger},\hat{\psi}_{j}$
and $\hat{c}_{n}^{\dagger},\hat{c}_{n}$ by
\begin{equation}
	\hat{\psi}_{j}^{\dagger}|\emptyset\rangle=|\psi_{j}\rangle=\sum_{n}|n\rangle\langle n|\psi_{j}\rangle=\sum_{n}a_{nj}\hat{c}_{n}^{\dagger}|\emptyset\rangle,\label{eq:PsiC1}
\end{equation}
\begin{equation}
	\hat{c}_{n}^{\dagger}|\emptyset\rangle=|n\rangle=\sum_{j}|\psi_{j}\rangle\langle\psi_{j}|n\rangle=\sum_{j}a_{nj}^{*}\hat{\psi}_{j}^{\dagger}|\emptyset\rangle,\label{eq:CPsi1}
\end{equation}
where $a_{nj}=\langle n|\psi_{j}\rangle$, or equivalently
\begin{equation}
	\hat{\psi}_{j}^{\dagger}=\sum_{n}a_{nj}\hat{c}_{n}^{\dagger},\qquad\hat{\psi}_{j}=\sum_{n}a_{nj}^{*}\hat{c}_{n},\label{eq:PsiC}
\end{equation}
\begin{equation}
	\hat{c}_{n}^{\dagger}=\sum_{j}a_{nj}^{*}\hat{\psi}_{j}^{\dagger},\qquad\hat{c}_{n}=\sum_{j}a_{nj}\hat{\psi}_{j}.\label{eq:CPsi}
\end{equation}
It can be verified that $\hat{\psi}_{j}$ and $\hat{\psi}_{j}^{\dagger}$
also satisfy the anticommutation relation of fermions, i.e.,
\begin{equation}
	\{\hat{\psi}_{i},\hat{\psi}_{j}\}=\{\hat{\psi}_{i}^{\dagger},\hat{\psi}_{j}^{\dagger}\}=0,\qquad\{\hat{\psi}_{i},\hat{\psi}_{j}^{\dagger}\}=\delta_{ij}.\label{eq:ACR2}
\end{equation}

We next construct the many-particle density operator for a set of
occupied Floquet states. Since the quasienergies $\{E_{j}\}$ of $\hat{U}$
are phase factors, they do not have a natural order even though they
are stroboscopically conserved in Floquet dynamics. Therefore, the
definition of a ``quasienergy Fermi surface'' below which each state
$|\psi_{j}\rangle$ is occupied by one fermion is ambiguous. Under
the conditions: (i) the quasienergies $\{E_{j}\}$ are confined to
the range of $[-\pi,\pi)$, (ii) the configuration of quasienergies is
symmetric with respect to $E=0$, and (iii) the quasienergy spectrum
is gapped at $E=0,\pm\pi$, it may look natural to define a ``Floquet
ground state'' by uniformly filling all states $\{|\psi_{j}\rangle\}$
with quasienergies ranging from $-\pi$ to zero. We make this choice
for most of the numerical calculations presented in this work. Here,
to be general enough, we just assume that a collection Floquet states
$\{|\psi_{j}\rangle|j\in{\rm occ.}\}$ are filled initially, such
that the many-particle wave function of the system takes the form
\begin{equation}
	|\Psi\rangle=\prod_{j\in{\rm occ.}}\hat{\psi}_{j}^{\dagger}|\emptyset\rangle,\label{eq:GS}
\end{equation}
which is normalized as $\langle\Psi|\Psi\rangle=1$ and
fulfills the eigenvalue equation $\hat{U}|\Psi\rangle=e^{-i\sum_{j\in{\rm occ.}}E_{j}}|\Psi\rangle$. 
The density operator corresponding to such a many-particle Floquet state
takes the form
\begin{equation}
	\hat{\rho}=|\Psi\rangle\langle\Psi|.\label{eq:R}
\end{equation}
It satisfies the general feature of a pure-state density matrix, i.e.,
$\hat{\rho}=\hat{\rho}^{\dagger}$, $\hat{\rho}=\hat{\rho}^{2}$ and
${\rm Tr}\hat{\rho}=1$.

To investigate the ES and EE, we decompose the system into two parts
A and B with Hilbert spaces ${\cal H}_{A}$ and ${\cal H}_{B}$, such
that the Hilbert space of the whose system ${\cal H}={\cal H}_{A}\otimes{\cal H}_{B}$.
For 1D and 2D systems, such a bipartition is illustrated in Fig.~\ref{fig:Sketch}.
Tracing over all degrees of freedom belonging to the subsystem B (${\rm Tr}_{{\rm B}}$)
yields the reduced density matrix $\hat{\rho}_{{\rm A}}$ of subsystem
A, i.e.,
\begin{equation}
	\hat{\rho}_{{\rm A}}={\rm Tr}_{{\rm B}}\hat{\rho}=\frac{1}{Z}e^{-\hat{H}_{{\rm A}}},\qquad Z\equiv{\rm Tr}e^{-\hat{H}_{{\rm A}}}.\label{eq:RA}
\end{equation}
In the second equality, we introduced the entanglement Hamiltonian
$\hat{H}_{{\rm A}}$, whose eigenvalues form the ES of the reduced
density matrix $\hat{\rho}_{{\rm A}}$. Note that even though the
Hamiltonian of the whole system $\hat{H}(t)$ is time-dependent, the
$\hat{H}_{{\rm A}}$ as constructed from Floquet eigenstates is time-independent
in stroboscopic dynamics. Due to the Hermiticity of $\hat{\rho}_{{\rm A}}$,
$\hat{H}_{{\rm A}}$ admits a spectral decomposition
\begin{equation}
	\hat{H}_{{\rm A}}=\sum_{j}\xi_{j}\hat{\phi}_{j}^{\dagger}\hat{\phi}_{j},\label{eq:HA}
\end{equation}
where $\hat{\phi}_{j}^{\dagger}$ creates an eigenstate $|\phi_{j}\rangle$
of $\hat{H}_{{\rm A}}$ with the entanglement eigenvalue $\xi_{j}\in\mathbb{R}$,
i.e., $|\phi_{j}\rangle=\hat{\phi}_{j}^{\dagger}|\emptyset\rangle$ and $\hat{H}_{{\rm A}}|\phi_{j}\rangle=\xi_{j}|\phi_{j}\rangle$.
The ES is formed by the collection of $\xi_j$ for all $j$.
Despite the ES, we will also study the EE of Floquet system. The von
Neumann EE for $\hat{\rho}_{{\rm A}}$ is defined as
\begin{equation}
	S=-{\rm Tr}\left(\hat{\rho}_{{\rm A}}\ln\hat{\rho}_{{\rm A}}\right).\label{eq:EE1}
\end{equation}
A larger value of $S$ indicates that the subsystems A and B are entangled
more strongly across their interface. Meanwhile, the configuration
of ES may provide more information about topological edge states in
the system even under the PBC.

For static noninteracting systems, due to the Wick's theorem, both the ES and EE can be obtained
from the spectrum of single-particle correlation matrix of the occupied
state $|\Psi\rangle$ restricted to the subsystem A \cite{EntangleRev4}.
A similar relation can be derived for the reduced density matrix of
Floquet states. To see this, we first express the matrix elements of
the single particle correlator as
\begin{equation}
	C_{mn}=\langle\Psi|\hat{c}_{m}^{\dagger}\hat{c}_{n}|\Psi\rangle,\label{eq:CM1}
\end{equation}
where the indices $m$ and $n$ are restricted to the subsystem A.
Using Eqs.~(\ref{eq:CPsi})--(\ref{eq:GS}),
$C_{mn}$ can be cast into the form
\begin{equation}
	C_{mn}=\sum_{j\in{\rm occ.}}\langle n|\psi_{j}\rangle\langle\psi_{j}|m\rangle=\langle n|\hat{P}|m\rangle,\label{eq:CM2}
\end{equation}
where $\hat{P}=\sum_{j\in{\rm occ.}}|\psi_{j}\rangle\langle\psi_{j}|$
is the single-particle projector onto the occupied Floquet states
in $|\Psi\rangle$. We can use Eq.~(\ref{eq:CM2}) to obtain the correlation
matrix in numerical calculations.

We next establish the relation between the spectrum of correlator $C$
and the entanglement Hamiltonian $\hat{H}_{{\rm A}}$. First, $C_{mn}$
can be equivalently expressed as
\begin{equation}
	C_{mn}={\rm Tr}\left(\hat{c}_{m}^{\dagger}\hat{c}_{n}\hat{\rho}_{{\rm A}}\right),\label{eq:CM3}
\end{equation}
since the operator $\hat{c}_{m}^{\dagger}\hat{c}_{n}$ belongs to
the subsystem A for all $m,n\in{\rm A}$. Plugging Eqs.~(\ref{eq:RA})
and (\ref{eq:HA}) into Eq.~(\ref{eq:CM3}), we obtain
\begin{equation}
	C_{mn}=\frac{1}{Z}{\rm Tr}\left(\hat{c}_{m}^{\dagger}\hat{c}_{n}e^{-\sum_{j}\xi_{j}\hat{\phi}_{j}^{\dagger}\hat{\phi}_{j}}\right).\label{eq:CM4}
\end{equation}
When restricted to the subsystem A, both $\{|n\rangle\}$ and $\{|\phi_{j}\rangle\}$
form complete bases of A. Therefore, similar to the Eq.~(\ref{eq:CPsi}),
we can express the relation between $\hat{c}_{n}^{\dagger},\hat{c}_{n}$
and $\hat{\phi}_{j}^{\dagger},\hat{\phi}_{j}$ as
\begin{equation}
	\hat{c}_{n}^{\dagger}=\sum_{j}b_{nj}^{*}\hat{\phi}_{j}^{\dagger},\qquad\hat{c}_{n}=\sum_{j}b_{nj}\hat{\phi}_{j},\label{eq:CPhi}
\end{equation}
where $b_{nj}=\langle n|\phi_{j}\rangle$. Inserting Eq.~(\ref{eq:CPhi})
into Eq.~(\ref{eq:CM4}) yields
\begin{equation}
	C_{mn}=\frac{1}{Z}\sum_{j',j''}b_{mj'}^{*}b_{nj''}{\rm Tr}\left(\hat{\phi}_{j'}^{\dagger}\hat{\phi}_{j''}e^{-\sum_{j}\xi_{j}\hat{\phi}_{j}^{\dagger}\hat{\phi}_{j}}\right),\label{eq:CM5}
\end{equation}
which will vanish once $j'\neq j''$. Thus we find
\begin{alignat}{1}
	C_{mn}= & \sum_{j'}b_{mj'}^{*}b_{nj'}\frac{1}{Z}{\rm Tr}\left(\hat{\phi}_{j'}^{\dagger}\hat{\phi}_{j'}e^{-\sum_{j}\xi_{j}\hat{\phi}_{j}^{\dagger}\hat{\phi}_{j}}\right)\nonumber \\
	= & \sum_{j'}b_{mj'}^{*}b_{nj'}\frac{1}{Z}\left(-\frac{\partial}{\partial\xi_{j'}}\right){\rm Tr}\left(e^{-\sum_{j}\xi_{j}\hat{\phi}_{j}^{\dagger}\hat{\phi}_{j}}\right).\label{eq:CM6}
\end{alignat}
To proceed, we notice from Eqs.~(\ref{eq:RA}), (\ref{eq:HA}) and
the occupation nature of fermions that
\begin{equation}
	Z={\rm Tr}\left(e^{-\sum_{j}\xi_{j}\hat{\phi}_{j}^{\dagger}\hat{\phi}_{j}}\right)=\prod_{j}\left(1+e^{-\xi_{j}}\right).\label{eq:Z}
\end{equation}
Therefore, we have $\ln Z=\sum_{j}\ln(1+e^{-\xi_{j}})$ and
\begin{alignat}{1}
	\left(-\frac{\partial}{\partial\xi_{j'}}\right)\ln Z= & \frac{1}{Z}\left(-\frac{\partial}{\partial\xi_{j'}}\right){\rm Tr}\left(e^{-\sum_{j}\xi_{j}\hat{\phi}_{j}^{\dagger}\hat{\phi}_{j}}\right)\nonumber \\
	= & \frac{1}{e^{\xi_{j'}}+1}.\label{eq:Key}
\end{alignat}
Comparing Eq.~(\ref{eq:Key}) and the second line of Eq.~(\ref{eq:CM6}),
we arrive at
\begin{equation}
	C_{mn}=\sum_{j}\frac{b_{mj}^{*}b_{nj}}{e^{\xi_{j}}+1}=\sum_{j}\frac{\langle n|\phi_{j}\rangle\langle\phi_{j}|m\rangle}{e^{\xi_{j}}+1}.\label{eq:CM7}
\end{equation}
The transpose of correlation matrix $C$ thus admits the spectral
decomposition
\begin{equation}
	C^{\top}=\sum_{j}\zeta_{j}|\phi_{j}\rangle\langle\phi_{j}|,\qquad\zeta_{j}=\frac{1}{e^{\xi_{j}}+1}.\label{eq:CM}
\end{equation}
Since $C$ and $C^{\top}$ share the same spectrum, we establish the
one-to-one correspondence between the spectrum $\{\xi_{j}\}$ of entanglement
Hamiltonian $H_{{\rm A}}$ and the spectrum $\{\zeta_{j}\}$ of correlation
matrix $C$, i.e.,
\begin{equation}
	\xi_{j}=\ln\left(\zeta_{j}^{-1}-1\right),\qquad j\in{\rm A}.\label{eq:ES}
\end{equation}
Note in passing that the ranges of eigenvalues are $\xi_{j}\in\mathbb{R}$ and $\zeta_{j}\in(0,1)$.
Eq.~(\ref{eq:ES}) allows us to determine the ES from the spectrum of single-particle correlation matrix for free fermions in Floquet systems.

The EE can also be extracted from the spectrum of correlation matrix.
Plugging Eqs.~(\ref{eq:RA}) and (\ref{eq:Z}) into Eq.~(\ref{eq:EE1}),
we find
\begin{alignat}{1}
	S= & -{\rm Tr}\left[\frac{1}{Z}e^{-H_{{\rm A}}}\ln\left(\frac{1}{Z}e^{-H_{{\rm A}}}\right)\right]\nonumber \\
	= & \sum_{j}\ln\left(1+e^{-\xi_{j}}\right)+\frac{1}{Z}{\rm Tr}\left(H_{{\rm A}}e^{-H_{{\rm A}}}\right).\label{eq:EE2}
\end{alignat}
With the help of Eqs.~(\ref{eq:HA}) and (\ref{eq:CM6})\textendash (\ref{eq:Key}),
we further obtain
\begin{alignat}{1}
	& {\rm Tr}\left(H_{{\rm A}}e^{-H_{{\rm A}}}\right)=\sum_{j}\xi_{j}{\rm Tr}\left(\hat{\phi}_{j}^{\dagger}\hat{\phi}_{j}e^{-H_{{\rm A}}}\right)\nonumber \\
	= & \sum_{j}\xi_{j}\left(-\frac{\partial}{\partial\xi_{j}}\right){\rm Tr}\left(e^{-H_{{\rm A}}}\right)=Z\sum_{j}\frac{\xi_{j}}{e^{\xi_{j}}+1}.\label{eq:Key2}
\end{alignat}
The expression for $S$ in Eq.~(\ref{eq:EE2}) can then be reduced
to
\begin{equation}
	S=\sum_{j}\left[\ln\left(1+e^{-\xi_{j}}\right)+\frac{\xi_{j}}{e^{\xi_{j}}+1}\right].\label{eq:EE3}
\end{equation}
Replacing $\xi_{j}$ by $\zeta_{j}$ following Eq.~(\ref{eq:ES}),
we arrive at the expression of EE in terms of the eigenvalues of correlation
matrix $C$, i.e.,
\begin{equation}
	S=-\sum_{j}\left[\zeta_{j}\ln\zeta_{j}+\left(1-\zeta_{j}\right)\ln\left(1-\zeta_{j}\right)\right].\label{eq:EE}
\end{equation}

In previous studies, the Eqs.~(\ref{eq:ES}) and (\ref{eq:EE}) for the
ES and EE have been derived for static systems of free fermions \cite{EntangleRev4}.
Here we showed that they can also be used to describe the ES and EE
of noninteracting fermions in Floquet systems for rather general partitions
of the system and populations of the initial Floquet state. In 
Sec.~\ref{sec:Res}, we will employ the connection between the correlation
matrix, ES and EE to further uncover the topological and entanglement
nature of typical Floquet systems in one and two spatial dimensions.

Besides the ES and EE, we can also characterize the nature of reduced density matrix (or entanglement Hamiltonian) of Floquet states through topological
invariants, whose definitions are insensitive to the choice of boundary
conditions. In one spatial dimension, if the Floquet operator of the
system possesses the chiral (sublattice) symmetry (i.e., for systems belonging to the symmetry classes AIII, BDI and CII), we can introduce
an open-bulk winding number \cite{OBWN1,OBWN2,OBWN3,OBWN4} for the
resulting correlation matrix $C$, which is defined as
\begin{equation}
	W=\frac{1}{L'}{\rm Tr}'(\Gamma_{{\rm A}}Q[Q,N_{{\rm A}}]).\label{eq:W}
\end{equation}
Here $\Gamma_{{\rm A}}$ and $N_{{\rm A}}$ are the chiral symmetry
operator and position operator of the subsystem A. In the definition,
the subsystem A is decomposed into a bulk region and two edge regions
around its left and right boundaries. The trace ${\rm Tr}'$ is taken
over the bulk region of A, whose size is $L'=L_{{\rm A}}-2L_{{\rm E}}$,
where $L_{{\rm A}}$ is the length of subsystem A and $L_{{\rm E}}$
counts the number of lattice sites belonging to the left and right
edge regions of A. The projector matrix $Q$ is defined as
\begin{equation}
	Q=\sum_{j}[\Theta(\zeta_{j}-1/2)-\Theta(1/2-\zeta_{j})]|\phi_{j}\rangle\langle\phi_{j}|,\label{eq:Q}
\end{equation}
where the step function $\Theta(\zeta)=1$ ($=0$) if $\zeta>0$ ($\zeta<0$).
$\{\text{\ensuremath{\zeta_{j}}}\}$ and $\{|\phi_{j}\rangle\}$ are
eigenvalues and eigenvectors of the correlation matrix $C$. In 
Sec.~\ref{subsec:ESEE1D}, we will demonstrate that when working in symmetric
time frames, the winding number $W$ introduced here could fully capture
the topological phases of 1D Floquet systems in the symmetry classes
BDI and CII.

In two spatial dimensions, we will focus on the ES and EE of Floquet
Chern insulators. For that purpose, we introduce the local Chern marker
(LCM) \cite{RSCN1,RSCN2,RSCN3} for the entanglement Hamiltonian, i.e.,
\begin{equation}
	\mathfrak{C}(x,y)=-4\pi{\rm Im}\langle x,y|P\hat{x}P\hat{y}P|x,y\rangle,\label{eq:LCM}
\end{equation}
where $\hat{x},\hat{y}$ are position operators along the $x,y$ directions,
and $\{|x,y\rangle\}$ refers to position eigenbasis of the 2D plane.
$P$ is the projector to the eigenstates of entanglement Hamiltonian
at half filling, i.e.,
\begin{equation}
	P=\sum_{j}\Theta(1/2-\zeta_{j})|\phi_{j}\rangle\langle\phi_{j}|.\label{eq:P}
\end{equation}
From $\mathfrak{C}(x,y)$, one can extract a real-space Chern number
that works equally well for uniform and non-uniform samples, which
is defined as
\begin{equation}
	{\rm Ch}=\frac{1}{L_{x}L_{y}}\sum_{x,y}\mathfrak{C}(x,y),\label{eq:CN}
\end{equation}
where the coordinates $(x,y)$ belong to a region of the subsystem
A with size $L_{x}\times L_{y}$ that is chosen away from the boundaries
of A. The real-space Chern number ${\rm Ch}$ has been used to characterize
the topology of Chern insulators in the presence of junctions and
various types of onsite potential \cite{RSCN1,RSCN2,RSCN3,RSCN4,RSCN5}. In
Sec.~\ref{subsec:ESEE2D}, we will demonstrate that the ${\rm Ch}$ in
Eq.~(\ref{eq:CN}) can also be used to describe the topological properties
of entanglement Hamiltonians for 2D Floquet systems.

\section{Results\label{sec:Res}}

In this section, we use the entanglement tools introduced in 
Sec.~\ref{sec:The} to study topological phases and phase transitions in
1D and 2D Floquet systems. In one dimension, we consider two prototypical
Floquet topological insulator models in the symmetry classes BDI and
CII, which are characterized by $\mathbb{Z}\times\mathbb{Z}$ and
$2\mathbb{Z}\times2\mathbb{Z}$ topological invariants. Under the
OBC, they possess twofold and fourfold degenerate edge modes at the
quasienergies zero and $\pi$. We discuss how the numbers of these
topological edge modes and the transition points where these numbers
change could be related to the ES and EE of the corresponding Floquet
systems under the PBC in Sec.~\ref{subsec:ESEE1D}. In two dimensions,
we demonstrate the connection between the topological property of
a uniformly filled Floquet band and the gapless chiral edge modes
in their related ES for both two- and three-band Floquet Chern insulator
models. The EE and the real-space Chern number of the entanglement
Hamiltonian are further shown to be able to capture the topological
phase transitions in these systems.

\subsection{ES and EE in 1D Floquet systems\label{subsec:ESEE1D}}

\subsubsection{Floquet topological insulators in BDI symmetry class\label{subsec:KRS}}

We first consider a model of Floquet topological insulator in the
symmetry class BDI, which is characterized by $\mathbb{Z}\times\mathbb{Z}$
topological invariants. It was first proposed as a spin-$1/2$ extension
of the on-resonance double kicked rotor (ORDKR) in Ref.~\cite{KRSZhou2018},
and later realized experimentally by an NV center in diamond in 
Ref.~\cite{KRSChen2021}. The possibility of realizing this model in cold atom
systems is also discussed in detail in Ref.~\cite{KRSBolik2022}.
The Floquet operator of the model takes the form
\begin{equation}
	\hat{U}=e^{-i\hat{H}_{2}}e^{-i\hat{H}_{1}},\label{eq:KRSU}
\end{equation}
where
\begin{equation}
	\hat{H}_{1}=\frac{K_{1}}{2}\sum_{n}(\hat{{\bf c}}_{n}^{\dagger}\sigma_{x}\hat{{\bf c}}_{n+1}+{\rm H.c.}),\label{eq:KRSH1}
\end{equation}
\begin{equation}
	\hat{H}_{2}=\frac{K_{2}}{2i}\sum_{n}(\hat{{\bf c}}_{n}^{\dagger}\sigma_{y}\hat{{\bf c}}_{n+1}-{\rm H.c.}).\label{eq:KRSH2}
\end{equation}
Here $K_{1}$ and $K_{2}$ are kicking strengths \cite{KRSZhou2018}.
$\sigma_{x}=\left|\uparrow\right\rangle\left\langle\downarrow\right|+\left|\downarrow\right\rangle\left\langle\uparrow\right|$
and $\sigma_{y}=i(\left|\downarrow\right\rangle\left\langle\uparrow\right|-\left|\uparrow\right\rangle\left\langle\downarrow\right|)$
are Pauli matrices acting on the spin-$1/2$ degree of freedom. $\hat{{\bf c}}_{n}\equiv(\hat{c}_{n,\uparrow},\hat{c}_{n,\downarrow})^{\top}$,
and $\hat{c}_{n,\sigma}$ is the annihilation operator of a fermion
with the momentum index $n\in\mathbb{Z}$ and spin $\sigma\in\{\uparrow,\downarrow\}$
\cite{KRSZhou2018}. $\top$ denotes the matrix transpose and ${\rm H.c.}$
means the Hermitian conjugate.
To describe the symmetry and topology
of the system, one can introduce symmetric time frames
by performing unitary transformations to $\hat{U}$, so that the transformed Floquet operators possess the chiral symmetry explicitly and a topological winding number can be defined for each of them appropriately~\cite{AsbothSTM}.
For 1D systems, there are two such time frames available in the usual case, and the Floquet operators in these time frames can be obtained by shifting the initial time of the evolution forward and backward over a quarter of the driving period. 
For our model, the resulting Floquet operators in a pair of complementary symmetric time frames read
\begin{equation}
	\hat{U}_{1}=e^{-i\frac{\hat{H}_{1}}{2}}e^{-i\hat{H}_{2}}e^{-i\frac{\hat{H}_{1}}{2}},\qquad\hat{U}_{2}=e^{-i\frac{\hat{H}_{2}}{2}}e^{-i\hat{H}_{1}}e^{-i\frac{\hat{H}_{2}}{2}}.\label{eq:KRSU12}
\end{equation}
In these time frames, the Floquet operators possess the chiral symmetry
$\Gamma=\sigma_{z}$ in the sense that $\Gamma\hat{U}_{\alpha}\Gamma^{-1}=\hat{U}_{\alpha}^{\dagger}$
($\alpha=1,2$). This implies that for any Floquet eigenstate $|\psi_{j}\rangle$
of $\hat{U}_{\alpha}$ with the quasienergy $E_{j}$ (i.e., $\hat{U}_{\alpha}|\psi_{j}\rangle=e^{-iE_{j}}|\psi_{j}\rangle$),
there must be another eigenstate $\Gamma|\psi_{j}\rangle$ of $\hat{U}_{\alpha}$
with the quasienergy $-E_{j}$. Floquet states with $E_{j}=0$ or $E_{j}=\pi$
then must come as degenerate pairs. The same spectral symmetry also holds
for the Floquet operator $\hat{U}$ in the original time frame [Eq.~(\ref{eq:KRSU})], as unitary transformations
do not change the Floquet spectrum obtained by solving $\hat{U}|\psi\rangle=e^{-iE}|\psi\rangle$ \cite{AsbothSTM}.

Besides the chiral symmetry, $\hat{U}_{\alpha}$ also possesses the
time-reversal symmetry ${\cal K}$ and particle-hole symmetry $\sigma_{z}{\cal K}$,
where ${\cal K}$ takes the complex conjugation.
It thus belongs to the symmetry class BDI \cite{AsbothSTM}. In 
Ref.~\cite{KRSZhou2018}, two integer-quantized topological winding numbers
$(w_{1},w_{2})$, defined for the $\hat{U}_{1}$ and $\hat{U}_{2}$ in symmetric time frames, were
used to characterize the Floquet topological phases and bulk-edge
correspondence of the spin-$1/2$ ORDKR. With the increase of kicking
strengths $(K_{1},K_{2})$, rich Floquet phases featured by large
winding numbers $(w_{0},w_{\pi})=(w_{1}+w_{2},w_{1}-w_{2})/2$ and
many topological edge states at zero and $\pi$ quasienergies can
be induced~\cite{KRSZhou2018}. Different Floquet topological phases
are separated by the phase transition lines
\begin{equation}
	\frac{\mu^{2}}{K_{1}^{2}}+\frac{\nu^{2}}{K_{2}^{2}}=\frac{1}{\pi^{2}},\qquad\mu,\nu\in\mathbb{Z}\label{eq:KRSPsBd}
\end{equation}
in the parameter space \cite{KRSZhou2018}. Across each phase boundary
line, $w_{0}$ or $w_{\pi}$ gets a quantized change. Within each topological
phase, the winding numbers $(w_{0},w_{\pi})$ are related to the number
of Floquet edge modes at zero and $\pi$ quasieneriges $(n_{0},n_{\pi})$
through the bulk-edge correspondence relation $(n_{0},n_{\pi})=2(|w_{0}|,|w_{\pi}|)$
\cite{KRSZhou2018} $w_0$ ($w_\pi$) thus counts the number of degenerate Floquet edge modes at the quasienergy zero ($\pi$).

In the following, we reveal the topology, phase transitions and bulk-edge
correspondence of the spin-$1/2$ ORDKR by investigating its ES, EE
and the winding numbers of entanglement Hamiltonian. Starting with
the Floquet operators $\hat{U}_{1}$ and $\hat{U}_{2}$ in Eq.~(\ref{eq:KRSU12}),
we consider a half-filled system where all Floquet states with quasienergies
$E\in(-\pi,0)$ are uniformly occupied. Following Eqs.~(\ref{eq:U})\textendash (\ref{eq:CM}),
the correlation matrix reads
\begin{equation}
	C_{\alpha}^{\top}=\sum_{j}\zeta_{j}^{\alpha}|\phi_{j}^{\alpha}\rangle\langle\phi_{j}^{\alpha}|,\label{eq:KRSCM}
\end{equation}
where $\alpha=1,2$ are indices of the two symmetric time frames.
The ES and EE are further obtained according to Eqs.~(\ref{eq:ES})
and (\ref{eq:EE}), i.e., for $\alpha=1,2$ we have
\begin{equation}
	\xi_{j}^{\alpha}=\ln\left(1/\zeta_{j}^{\alpha}-1\right),\qquad j\in{\rm A},\label{eq:KRSES}
\end{equation}
\begin{equation}
	S_{\alpha}=-\sum_{j}\left[\zeta_{j}^{\alpha}\ln\zeta_{j}^{\alpha}+\left(1-\zeta_{j}^{\alpha}\right)\ln\left(1-\zeta_{j}^{\alpha}\right)\right].\label{eq:KRSEE}
\end{equation}
Meanwhile, we find the Floquet spectrum $E$ of $\hat{U}$ in 
Eq.~(\ref{eq:KRSU}) by solving the eigenvalue equation $\hat{U}|\psi\rangle=e^{-iE}|\psi\rangle$.
All calculations are performed under the PBC by setting $\hat{{\bf c}}_{n}=\hat{{\bf c}}_{n+L}$
in Eqs.~(\ref{eq:KRSH1}) and (\ref{eq:KRSH2}), with $L$ being the
length of lattice. A direct advantage of the entanglement approach
deserves to be mentioned. For the spin-$1/2$ ORDKR, the lattice structures
in Eqs.~(\ref{eq:KRSH1}) and (\ref{eq:KRSH2}) are defined in momentum
space \cite{KRSZhou2018}, where the realization of an OBC for detecting
the edge states there is practically challenging. As will be seen
in the following, the entanglement approach allows us to get access
to the information about Floquet edge states even under the PBC, which
can be implemented more naturally for the momentum space of a rotor.

\begin{figure}
	\begin{centering}
		\includegraphics[scale=0.49]{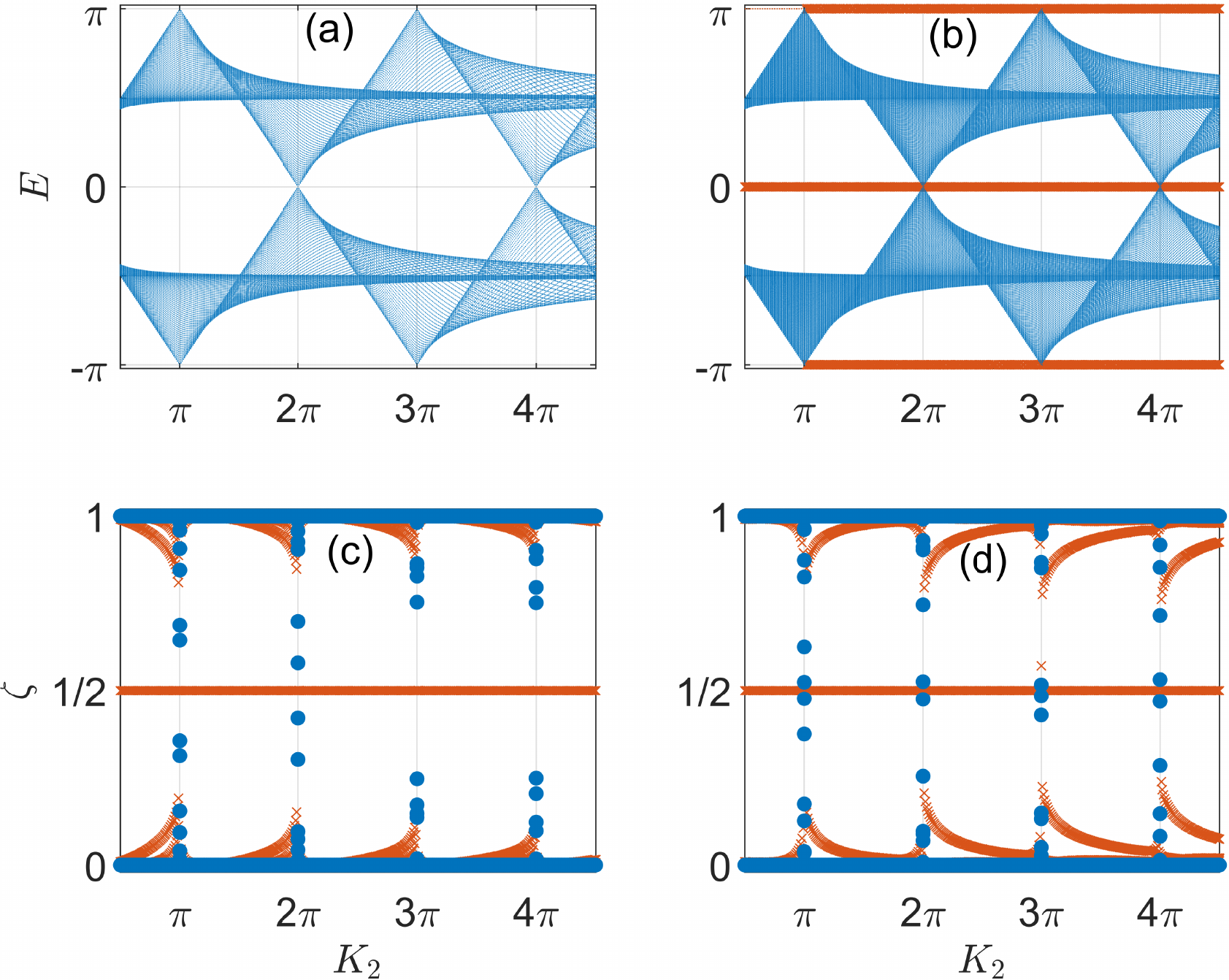}
		\par\end{centering}
	\caption{Quasienergy and correlation matrix spectra of the spin-$1/2$ ORDKR
		versus the kicking strength $K_{2}$. We set the kicking strength at
		$K_{1}=0.5\pi$. The length of lattice is $L=300$ for all panels.
		(a) shows the Floquet spectrum under the PBC. (b) shows the Floquet
		spectrum under the OBC, where the red crosses denote the edge states with
		quasienergies $E=0,\pm\pi$. (c) and (d) show the spectrum of correlation
		matrices $C_{1}$ and $C_{2}$ in two symmetric time frames. Red crosses
		in (c) and (d) denote eigenstates of $C_{1}$ and $C_{2}$ whose profiles
		are localized around the cuts between subsystems A and B. \label{fig:KRS-E-ES}}
\end{figure}

In Fig.~\ref{fig:KRS-E-ES}, we report the quasienergy spectrum and
ES (presented by the spectrum of correlation matrices in symmetric
time frames $\alpha=1,2$) of the spin-$1/2$ ORDKR versus the kicking
strength $K_{2}$. In Fig.~\ref{fig:KRS-E-ES}(a), we see that the
gap of Floquet spectrum closes/reopens alternately at $E=\pi$ and
$E=0$ for $K_{2}=\nu\pi$ ($\nu=1,2,...$), as predicted by Eq.~(\ref{eq:KRSPsBd}).
New Floquet edge states emerge at the quasienergy $E=0$ or $\pi$
after each transition, as shown in Fig.~\ref{fig:KRS-E-ES}(b). With
the increase of $K_{2}$, the number of zero ($\pi$) Floquet edge
modes changes in the sequence $n_{0}=2\rightarrow6\rightarrow10$
($n_{\pi}=0\rightarrow4\rightarrow8$) following each transition at
$E=0$ ($E=\pi$) \cite{KRSZhou2018}.

\begin{figure}
	\begin{centering}
		\includegraphics[scale=0.49]{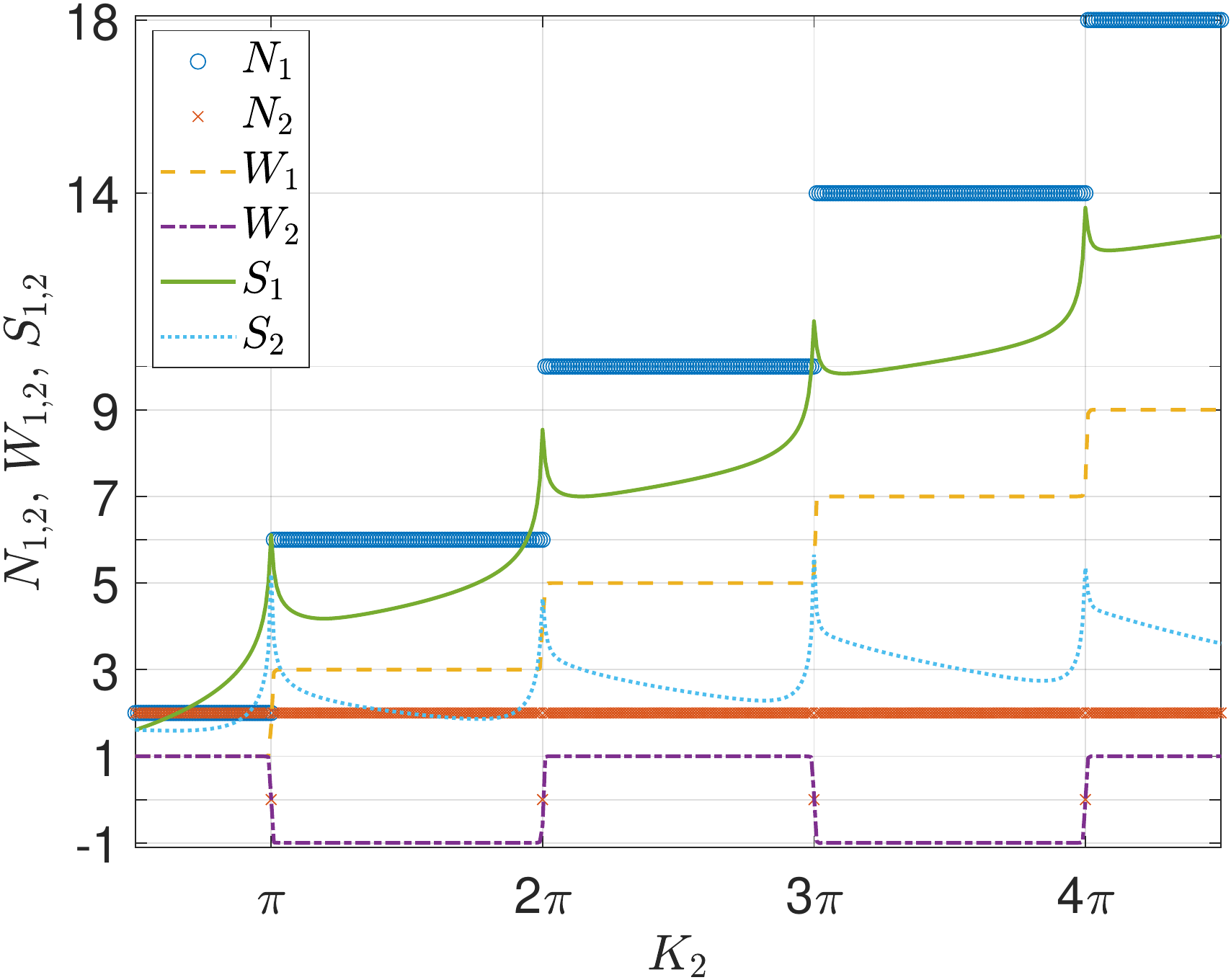}
		\par\end{centering}
	\caption{The number of maximally entangled states, winding numbers of the entanglement
		Hamiltonian and the EE evaluated in symmetric time frames $\alpha=1,2$.
		The kicking strength $K_{1}$ is set to $0.5\pi$ and the length of
		lattice is $L=300$. $N_{\alpha}$ denotes the number of eigenstates
		of the correlation matrix $C_{\alpha}$ whose eigenvalues are equal
		to $1/2$. $W_{\alpha}$ denotes the open-bulk winding number of the
		entanglement Hamiltonian {[}Eq.~(\ref{eq:HA}){]}. $S_{\alpha}$ denotes
		the EE {[}Eq.~(\ref{eq:KRSEE}){]} in the time frame $\alpha$. \label{fig:KRS-N-W-S}}
\end{figure}

Near the critical kicking strengths $K_{2}=\nu\pi$ ($\nu=1,2,...$),
the ES in Figs.~\ref{fig:KRS-E-ES}(c) and \ref{fig:KRS-E-ES}(d)
show discontinuous changes around the correlation matrix eigenvalue
$\zeta=1/2$. In Fig.~\ref{fig:KRS-N-W-S}, the EE in both symmetric
time frames also show nonanalytic cusps around $K_{2}=\nu\pi$, implying
the presence of phase transitions around these critical kicking strengths.
Referring to Figs.~\ref{fig:KRS-E-ES}(a) and (b), we confirm that
the ES and EE indeed provide signatures for Floquet topological phase
transitions in the spin-$1/2$ ORDKR.

Away from these transition points, the eigenvalues of $C_{\alpha}$
appear around $0$, $1/2$ and $1$. According to the Eq.~(\ref{eq:KRSEE}), the
eigenvalues $\zeta_{j}^{\alpha}\simeq0,1$ have vanishing contributions
to $S_{\alpha}$. Meanwhile, each eigenvalue $\zeta_{j}^{\alpha}=1/2$
has the largest contribution $\ln2$ to the EE in Fig.~\ref{fig:KRS-N-W-S}.
The corresponding eigenstate $|\phi_{j}^{\alpha}\rangle$ should be
maximally entangled, which means that it should appear around the
spatial cuts between the subsystems A and B. We thus refer to
the eigenstates $|\phi_{j}^{\alpha}\rangle$ of $C_{\alpha}$ with
eigenvalues $\zeta_{j}^{\alpha}=1/2$ as ``edge states'' of the
entanglement Hamiltonian. In Fig.~\ref{fig:KRS-N-W-S}, we also observe
a steady growth in the global profile of $S_{1}$, which implies an
increase of the number $N_1$ of entanglement edge states with $\zeta_{j}^{1}=1/2$ in the ES of $\hat{U}_{1}$.
Note in passing that the global profile of $S_2$ also shows a slow growth 
with the increase of $K_2$. Since we have $N_2=2$ at all $K_2\neq\nu\pi$, the $\zeta_{j}^{2}=1/2$ modes cannot contribute
and this increase should be due to the appearance of eigenmodes whose ES deviates further from $\zeta=0,1$ but not reaching $1/2$, 
as observed in Fig.~\ref{fig:KRS-E-ES}(d).

To further decode the topological nature of entanglement Hamiltonian,
we report the numbers of localized modes $(N_{1},N_{2})$ with eigenvalues
$(\zeta_{j}^{1},\zeta_{j}^{2})=(1/2,1/2)$ in the ES {[}Figs.~\ref{fig:KRS-E-ES}(c)\textendash (d){]}
and the winding numbers $(W_{1},W_{2})$ with respect to the kicking
strength $K_{2}$. $(W_{1},W_{2})$ are defined following Eqs.~(\ref{eq:W})
and (\ref{eq:Q}). For the spin-$1/2$ ORDKR, we have $\Gamma_{{\rm A}}=\Gamma=\sigma_{z}$
and the projector in Eq.~(\ref{eq:Q}) takes the form
\begin{equation}
	Q_{\alpha}=\sum_{j}[\Theta(\zeta_{j}^{\alpha}-1/2)-\Theta(1/2-\zeta_{j}^{\alpha})]|\phi_{j}^{\alpha}\rangle\langle\phi_{j}^{\alpha}|,\label{eq:KRSQ}
\end{equation}
where $\{\zeta_{j}^{\alpha}\}$ and $\{|\phi_{j}^{\alpha}\rangle\}$
are the eigenvalues and eigenbasis of the correlation matrices 
{[}Eq.~(\ref{eq:KRSCM}){]} in the time frames $\alpha=1,2$. The resulting
topological winding numbers {[}Eq.~(\ref{eq:W}){]} of the entanglement
Hamiltonian reads
\begin{equation}
	W_{\alpha}=\frac{1}{L'}{\rm Tr}'(\sigma_{z}Q_{\alpha}[Q_{\alpha},N_{{\rm A}}]),\quad\alpha=1,2.\label{eq:KRSW}
\end{equation}

We now establish the correspondence between the ES and Floquet topological
phases. First, by comparing $(N_{1},N_{2})$ in the ES of Fig.~\ref{fig:KRS-N-W-S}
and the number of edge modes $(n_{0},n_{\pi})$ with quasienergies
zero and $\pi$ in Fig.~\ref{fig:KRS-E-ES}(b), we arrive at the relations
\begin{equation}
	N_{1}=n_{0}+n_{\pi},\qquad N_{2}=|n_{0}-n_{\pi}|.\label{eq:KRSNn}
\end{equation}
That is, in the symmetric time frame $1$ ($2$), the number of modes
with eigenvalue $\zeta=1/2$ in the ES under the PBC is equal to the
sum of (the absolute difference between) the number of zero and $\pi$
Floquet edge modes in the quasienergy spectrum under the OBC. Notably,
the total number of Floquet topological edge modes $n_{0}+n_{\pi}$
in the quasienergy spectrum can be captured by the number of entanglement
eigenmodes $N_{1}$ in a single time frame. This indicates that the
ES of a Floquet system indeed contain rich information about its nonequilibrium
topological properties.

Second, according to Fig.~\ref{fig:KRS-N-W-S}, the $(N_{1},N_{2})$
and winding numbers $(W_{1},W_{2})$ are related by the relation
\begin{equation}
	N_{1}=2|W_{1}|,\qquad N_{2}=2|W_{2}|.\label{eq:KRSNw}
\end{equation}
Though obtained under the PBC, it may be viewed as a ``bulk-edge
correspondence'', as $N_{\alpha}$ ($\alpha=1,2$) counts the number
of modes with degenerate eigenvalue $\zeta=1/2$ that are localized
at the spatial entanglement cuts between subsystems A and B. Finally,
combining Eqs.~(\ref{eq:KRSNn}) and (\ref{eq:KRSNw}), we find the
relation between the topology of entanglement Hamiltonian under the
PBC and the Floquet zero/$\pi$ edge modes under the OBC, i.e., 
\begin{alignat}{1}
	n_{0}= & \frac{1}{2}|{\rm sgn}(W_{1})N_{1}+{\rm sgn}(W_{2})N_{2}|=|W_1+W_2|,\nonumber \\
	n_{\pi}= & \frac{1}{2}|{\rm sgn}(W_{1})N_{1}-{\rm sgn}(W_{2})N_{2}|=|W_1-W_2|.\label{eq:BBC1}
\end{alignat}

Eq.~(\ref{eq:BBC1}) allows us to determine the numbers of Floquet
topological edge modes in an arbitrary time frame under the OBC from
the entanglement information of the system under the PBC. We can thus
regard it as an entanglement bulk-edge correspondence for Floquet
systems. We have also checked the Eqs.~(\ref{eq:KRSNn})\textendash (\ref{eq:BBC1})
in other parameter regions of the spin-$1/2$ ORDKR and confirmed
their validness, which implies that they are generic for 1D Floquet
topological insulators in the BDI symmetry class. The Eqs.~(\ref{eq:KRSNn})\textendash (\ref{eq:BBC1})
thus form one set of key results in this work. The generality of this
topology-entanglement connection will be further demonstrated in the
next subsection by considering Floquet topological insulators in another
symmetry class.

\subsubsection{Floquet topological insulators in CII symmetry class\label{subsec:SCL}}

\begin{figure}
	\begin{centering}
		\includegraphics[scale=0.49]{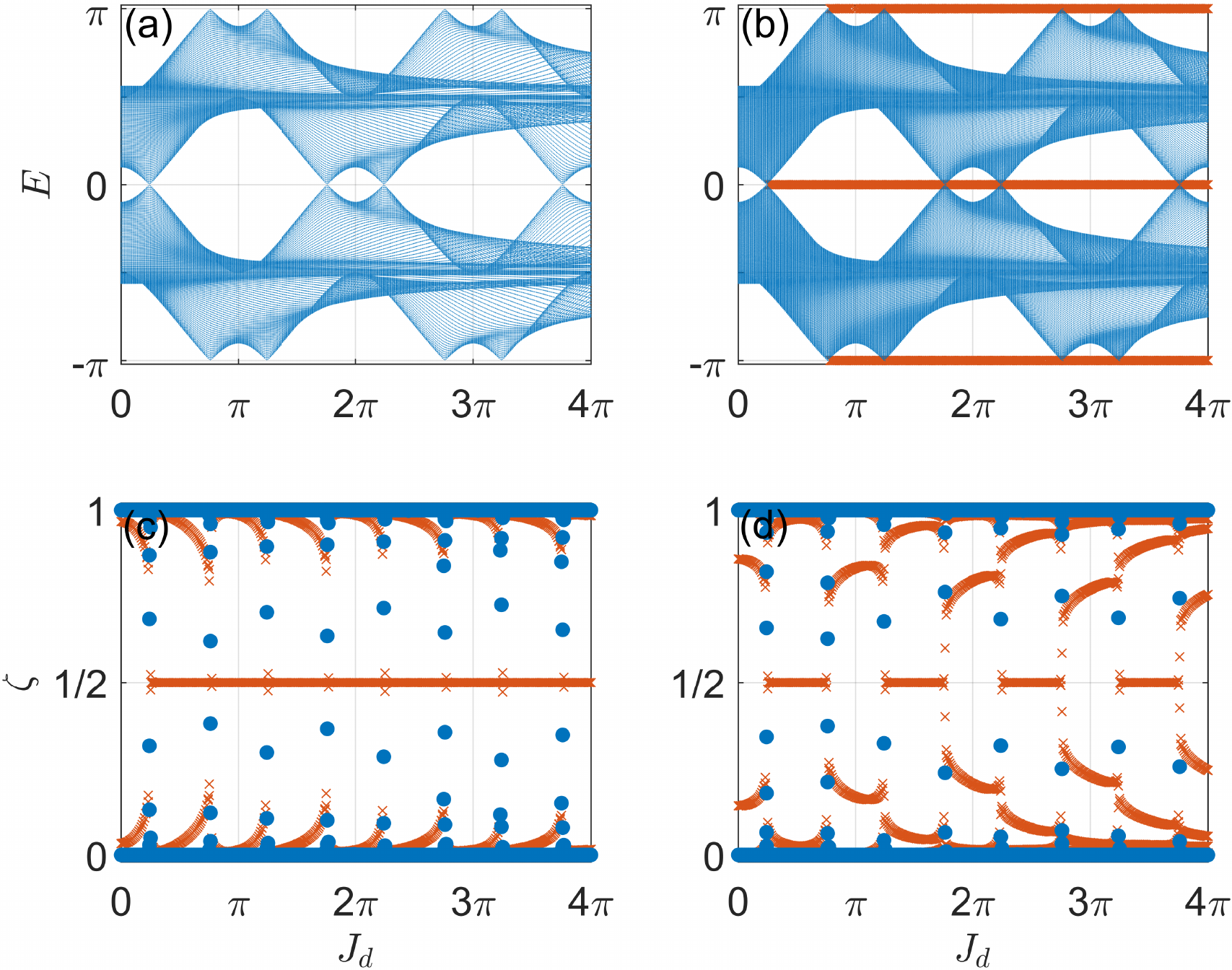}
		\par\end{centering}
	\caption{Quasienergy and correlation matrix spectrum of the spin-$1/2$ PQL
		versus the diagonal hopping amplitude $J_{d}$. Other system parameters
		are set as $(J_{x},J_{y},V)=(0.5\pi,0.6\pi,0.2\pi)$, and the length
		of lattice is $L=250$ for all panels. (a) shows the Floquet spectrum
		under the PBC. (b) shows the Floquet spectrum under the OBC, where
		the red crosses denote edge states with quasienergies $E=0,\pm\pi$.
		(c) and (d) show the spectrum of correlation matrices $C_{\alpha}$
		in symmetric time frames $\alpha=1,2$. Red crosses in (c) and (d)
		highlight eigenstates of $C_{\alpha}$ for $\alpha=1,2$ with localized
		profiles around the cuts between subsystems A and B. \label{fig:SCL-E-ES}}
\end{figure}

We now study a spin-$1/2$ periodically quenched ladder (PQL) model
\cite{SCLZhou2020}, whose Hamiltonian takes the form $\hat{{\cal H}}(t)=\hat{{\cal H}}_{1}$
in the time duration $t\in[\ell,\ell+1/2)$ and $\hat{{\cal H}}(t)=\hat{{\cal H}}_{2}$
in the time duration $t\in[\ell+1/2,\ell+1)$, where $\ell\in\mathbb{Z}$
and 
\begin{alignat}{1}
	\hat{{\cal H}}_{1}= & \sum_{n}J_{x}(\hat{{\bf c}}_{n}^{\dagger}\sigma_{0}\otimes\tau_{z}\hat{{\bf c}}_{n+1}+{\rm H.c.})\nonumber \\
	- & \sum_{n}iV(\hat{{\bf c}}_{n}^{\dagger}\sigma_{y}\otimes\tau_{0}\hat{{\bf c}}_{n+1}-{\rm H.c.}),\label{eq:SCLH1}
\end{alignat}
\begin{alignat}{1}
	\hat{{\cal H}}_{2}= & \sum_{n}J_{y}\hat{{\bf c}}_{n}^{\dagger}\sigma_{0}\otimes\tau_{x}\hat{{\bf c}}_{n}\nonumber \\
	+ & \sum_{n}iJ_{d}(\hat{{\bf c}}_{n}^{\dagger}\sigma_{z}\otimes\tau_{x}\hat{{\bf c}}_{n+1}-{\rm H.c.}).\label{eq:SCLH2}
\end{alignat}
Here have we set the driving period $T=1$. $J_{x}$, $J_{y}$ and $J_{d}$
denote the hopping amplitudes along the $x$, $y$, and diagonal directions
of the ladder. $V$ denotes the spin-orbit coupling between fermions
with opposite spins in adjacent unit cells \cite{SCLZhou2020}. $\sigma_{x,y,z}$
and $\tau_{x,y,z}$ are Pauli matrices acting on the spin-$1/2$ degree
of freedom and the two sublattices $(a,b)$, respectively. $\sigma_{0}$
and $\tau_{0}$ are $2\times2$ identity matrices. The vector annihilation
operator $\hat{{\bf c}}_{n}\equiv(\hat{c}_{n,\uparrow,a},\hat{c}_{n,\uparrow,b},\hat{c}_{n,\downarrow,a},\hat{c}_{n,\downarrow,b})^{\top}$,
where $\hat{c}_{n,\sigma,\tau}$ annihilates a fermion with spin $\sigma$
($=\uparrow,\downarrow$) at the sublattice $\tau$ ($=a,b$) in the
$n$th unit cell. The Floquet operator of the system takes the form
\begin{equation}
	\hat{{\cal U}}=e^{-\frac{i}{2}\hat{{\cal H}}_{2}}e^{-\frac{i}{2}\hat{{\cal H}}_{1}}.\label{eq:SCLU}
\end{equation}
To characterize the symmetry and topology of the system, we also transform
$\hat{{\cal U}}$ into symmetric time frames in parallel with what we did in Sec.~\ref{subsec:KRS}, yielding the Floquet operators
\begin{equation}
	\hat{{\cal U}}_{1}=e^{-\frac{i}{4}\hat{{\cal H}}_{1}}e^{-\frac{i}{2}\hat{{\cal H}}_{2}}e^{-\frac{i}{4}\hat{{\cal H}}_{1}},\quad\hat{{\cal U}}_{2}=e^{-\frac{i}{4}\hat{{\cal H}}_{2}}e^{-\frac{i}{2}\hat{{\cal H}}_{1}}e^{-\frac{i}{4}\hat{{\cal H}}_{2}}.\label{eq:SCLU12}
\end{equation}
$\hat{{\cal U}}_{1}$ and $\hat{{\cal U}}_{2}$ have been shown to
possess the time-reversal symmetry $i\sigma_{y}\otimes\tau_{0}{\cal K}$,
particle-hole symmetry $\sigma_{x}\otimes\tau_{y}{\cal K}$, and chiral symmetry
$\Gamma=-\sigma_{z}\otimes\tau_{y}$ with $\Gamma^{2}=1$ \cite{SCLZhou2020}.
As the time-reversal and particle-hole symmetries both square to $-1$,
the spin-$1/2$ PQL belongs to the symmetry class CII. The Floquet
topological phases of $\hat{{\cal U}}$ are thus characterized by
a pair of even-integer quantized winding numbers \cite{SCLZhou2020}.
In Figs.~\ref{fig:SCL-E-ES}(a) and \ref{fig:SCL-E-ES}(b), we show
the quasienergy spectrum of the system by solving the eigenvalue equation
$\hat{{\cal U}}|\psi\rangle=e^{-iE}|\psi\rangle$ under the PBC and
OBC, respectively. With the increase of $J_{d}$, a series of gap
closing/reopening processes are found at $E=0$ and $E=\pm\pi$. Fourfold
degenerate edge modes are observed at the quasienergies zero and $\pi$
(denoted by red stars). Both of their numbers $(n_{0},n_{\pi})$ change
in the sequence $0\rightarrow4\rightarrow8\rightarrow12\rightarrow16$
following the topological phase transitions induced by the increase
of $J_{d}$ \cite{SCLZhou2020}.

We now characterize the topology and phase transitions in this spin-$1/2$
PQL from the quantum entanglement perspective. We again focus on the
case with all Floquet states whose quasienergies $E\in(-\pi,0)$ are
uniformly filled. The spin-$1/2$ PQL possesses four quasienergy bands,
and the density operator of our consideration then refers to the occupation
of the two lower Floquet bands in the first quasienergy Brillouin
zone. In the $\alpha$'s time frame, the correlation matrix, ES and
EE of the spin-$1/2$ PQL share the same form with the spin-$1/2$
ORDKR, as described by Eqs.~(\ref{eq:KRSCM})\textendash (\ref{eq:KRSEE}).
In Figs.~\ref{fig:SCL-E-ES}(c) and \ref{fig:SCL-E-ES}(d), we show
the spectrum of correlation matrices $C_{1}$ and $C_{2}$ versus
$J_{d}$ under the PBC. Other parameters of $\hat{{\cal U}}_{1}$
and $\hat{{\cal U}}_{2}$ are chosen to be the same as those used
in Figs.~\ref{fig:SCL-E-ES}(a) and \ref{fig:SCL-E-ES}(b). We again
observe discontinuous behaviors in the ES whenever the quasienergy
spectrum gap of the spin-$1/2$ PQL closes at $E=0$ or $E=\pm\pi$.
Meanwhile, the numbers of localized modes in the ES with eigenvalue
$\zeta_{j}^{\alpha}=1/2$ change following different patterns in the
two time frames $\alpha=1$ and $2$. This situation is similar to
what we encountered in the spin-$1/2$ ORDKR. It implies that the
entanglement characteristics of Floquet reduced density matrices in
both two symmetric time frames are needed in order to capture the
full topological information of the system. We regard this as a generic
feature of 1D Floquet topological phases protected by chiral symmetry.

\begin{figure}
	\begin{centering}
		\includegraphics[scale=0.49]{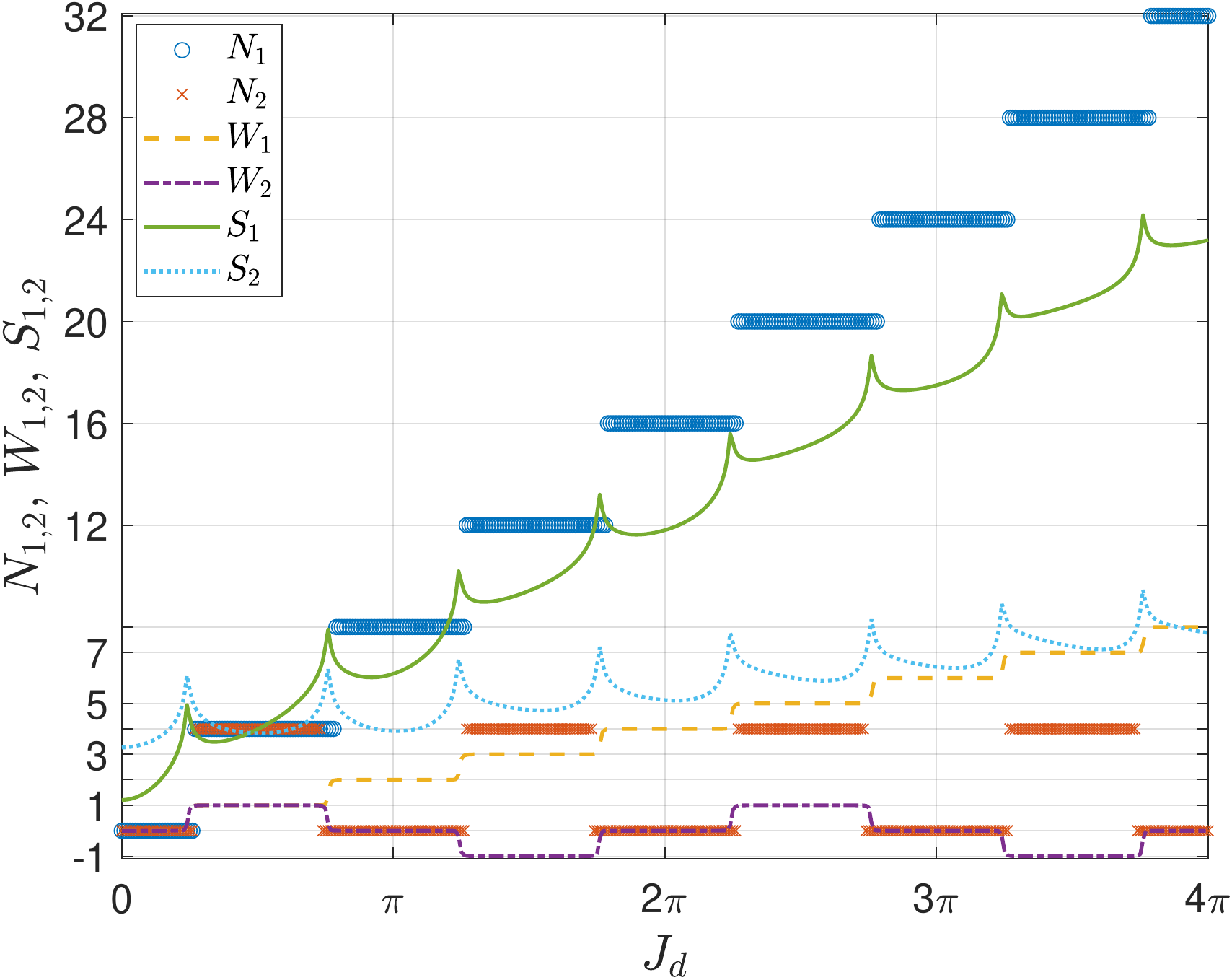}
		\par\end{centering}
	\caption{The number of the maximally entangled states, winding numbers of the entanglement
		Hamiltonian and the EE of the spin-$1/2$ PQL, evaluated in symmetric
		time frames $\alpha=1,2$. Other system parameters are $(J_{x},J_{y},V)=(0.5\pi,0.6\pi,0.2\pi)$,
		and the length of lattice is $L=250$. $N_{\alpha}$ denotes the number
		of eigenstates of the correlation matrix $C_{\alpha}$ with eigenvalue
		$1/2$. $W_{\alpha}$ is the open-bulk winding number of $C_{\alpha}$
		{[}Eq.~(\ref{eq:KRSCM}){]}. $S_{\alpha}$ denotes the EE in the time
		frame $\alpha$ {[}Eq.~(\ref{eq:KRSEE}){]}. \label{fig:SCL-N-W-S}}
\end{figure}

In Fig.~\ref{fig:SCL-N-W-S}, we first notice that in both time frames,
the EE shows a non-analytic cusp right at each gap closing point in
Figs.~\ref{fig:SCL-E-ES}(a) and \ref{fig:SCL-E-ES}(b). Therefore,
the EE can also be used to signify transitions between different Floquet
topological phases of the spin-$1/2$ PQL. Moreover, a steady growth
of $S_{1}$ is observed with the increase of $J_{d}$. It goes in
parallel with the monotonic raise of the number of $\zeta_{j}^{1}=1/2$ eigenmodes
by four after each transition. Every such eigenmode makes a largest
contribution $\ln2$ to $S_{1}$. 
Meanwhile, the $S_{2}$ also shows
a growth with a much smaller rate in its global profile. Since
the number of $\zeta_{j}^{2}=1/2$ eigenmodes $N_{2}$ only oscillates
between zero and four following the transitions induced by $J_{d}$,
instead of showing a monotonic increase as observed in $N_{1}$, 
the global increase in $S_2$ cannot be originated from these maximally entanglement eigenmodes.
Then there must be some other states whose ES deviates from zero and one further with the increase of $J_d$,
as we indeed observe in Fig.~\ref{fig:SCL-E-ES}(d).

To relate the topology of Floquet entanglement Hamiltonian under the
PBC and $\zeta=1/2$ eigenmodes with Floquet edge states under the
OBC, we construct the projector of correlation matrix eigenmodes in
the $\alpha$'s time frame as the form in Eq.~(\ref{eq:KRSQ}). The
open-bulk winding number {[}Eq.~(\ref{eq:W}){]} of entanglement Hamiltonian
in the $\alpha$'s time frame is then given by
\begin{equation}
	W_{\alpha}=\frac{1}{L'}{\rm Tr}'(\Gamma Q_{\alpha}[Q_{\alpha},N_{{\rm A}}]),\label{eq:PQLW}
\end{equation}
where $\alpha=1,2$. $\Gamma=-\sigma_{z}\otimes\tau_{y}$ is the chiral
symmetry operator of the spin-$1/2$ PQL. Other quantities share the
same meanings with those in Eq.~(\ref{eq:KRSW}). Referring to the
results reported in Figs.~\ref{fig:SCL-E-ES} and \ref{fig:SCL-N-W-S},
we conclude that the relation between the numbers
of $\zeta=1/2$ eigenmodes $(N_{1},N_{2})$ in the ES and the numbers
of Floquet edge modes $(n_{0},n_{\pi})$ with quasienergies zero and
$\pi$ in Eq.~(\ref{eq:KRSNn}) holds also for the spin-$1/2$ PQL. This observation implies
the generality of Eq.~(\ref{eq:KRSNn}) in describing the relationship
between entanglement and topological edge states for 1D Floquet topological
insulators in the symmetry class CII. Next, for the spin-$1/2$ PQL,
the relation between the numbers of $\zeta=1/2$
eigenmodes $(N_{1},N_{2})$ and the winding numbers $(W_{1},W_{2})$
in Eq.~(\ref{eq:KRSNw}) now takes the form
\begin{equation}
	N_{1}=4|W_{1}|,\qquad N_{2}=4|W_{2}|,\label{eq:PQLNw}
\end{equation}
as verified by the data reported in Fig.~\ref{fig:SCL-N-W-S}. 
The prefactor $4$ comes from the fact that the $\zeta=1/2$ eigenmodes are fourfold degenerate.
Eq.~(\ref{eq:PQLNw}) describes the bulk-edge correspondence of ES in the two symmetric
time frames. It is generic for CII class Floquet topological
phases in one-dimension. Finally, the combination of Eqs.~(\ref{eq:KRSNn})
and (\ref{eq:PQLNw}) yields the relation between the topology of
Floquet entanglement Hamiltonian under the PBC and the zero/$\pi$
Floquet edge modes of the spin-$1/2$ PQL under the OBC following
Eq.~(\ref{eq:BBC1}), or explicitly expressed as
\begin{equation}
	n_{0}=2|W_{1}+W_{2}|,\qquad n_{\pi}=2|W_{1}-W_{2}|.\label{eq:BBC2}
\end{equation}

Eq.~(\ref{eq:BBC2}) allows us to know the exact numbers of Floquet
zero and $\pi$ quasienergy edge modes under the OBC from the topology of ES under the PBC.
We regard it as an entanglement bulk-edge correspondence for 1D Floquet
systems in the CII symmetry class. We have tested these equations
in other parameter regions of the spin-$1/2$ PQL and confirmed their
correctness. They are therefore general for CII class 1D Floquet topological
matter. Eqs.~(\ref{eq:KRSNn}), (\ref{eq:PQLNw}) and (\ref{eq:BBC2})
form another set of key results in this work. In the next subsection,
we discuss the universality of the topology-entanglement connection in
2D Floquet topological insulators.

\subsection{ES and EE in 2D Floquet systems\label{subsec:ESEE2D}}

\subsubsection{Periodically quenched Chern insulator\label{subsec:GHM}}

We now investigate 2D Floquet Chern insulators, which could possess
topological phases with large Chern numbers. We start with spinless
noninteracting fermions in a periodically quenched generalized Haldane
model (PQGHM) with third neighbor hopping \cite{FloBigTN7,Haldane1,Haldane2,Haldane3}.
In the absence of quenches and under the PBC along two spatial dimensions,
the momentum-space Hamiltonian of the system takes the form $\hat{H}=\sum_{{\bf k}\in{\rm BZ}}\hat{{\bf c}}_{{\bf k}}^{\dagger}H({\bf k})\hat{{\bf c}}_{{\bf k}}$.
Here ${\bf k}=(k_{1},k_{2})\in[-\pi,\pi)\times[-\pi,\pi)$ is the
quasimomentum. $\hat{{\bf c}}_{{\bf k}}\equiv(\hat{c}_{{\bf k},a},\hat{c}_{{\bf k},b})^{\top}$,
where $\hat{c}_{{\bf k},a}$ ($\hat{c}_{{\bf k},b}$) annihilates
a fermion with quasimomentum ${\bf k}$ on the sublattice $a$ ($b$)
of the honeycomb lattice. The Bloch Hamiltonian is $H({\bf k})=h_{{\bf k}}^{x}\sigma_{x}+h_{{\bf k}}^{y}\sigma_{y}+h_{{\bf k}}^{z}\sigma_{z}$,
where $h_{{\bf k}}^{x}=t_{1}(1+\cos k_{1}+\cos k_{2})+t_{3}[2\cos(k_{1}-k_{2})+\cos(k_{1}+k_{2})]$,
$h_{{\bf k}}^{y}=t_{1}(\sin k_{1}+\sin k_{2})+t_{3}\sin(k_{1}+k_{2})$,
and $h_{{\bf k}}^{z}=2t_{2}\sin\phi[\sin k_{1}-\sin k_{2}-\sin(k_{1}-k_{2})]$
\cite{FloBigTN7}. The Pauli matrices $\sigma_{x,y,z}$ act on the
sublattice degrees of freedom. The quench is applied by decomposing
each driving period $T$ into two episodes $T_{1}$ and $T_{2}$ ($T=T_{1}+T_{2}$),
and setting the parameters $(t_{3},\phi)=(t_{31},\phi_{1})$ ($(t_{3},\phi)=(t_{32},\phi_{2})$)
in the time duration $T_{1}$ ($T_{2}$). We denote the system's Bloch
Hamiltonians during $T_{1}$ and $T_{2}$ as $H_{1}({\bf k})$ and
$H_{2}({\bf k})$. The Floquet operator then takes the form $\hat{U}=\sum_{{\bf k}\in{\rm BZ}}\hat{{\bf c}}_{{\bf k}}^{\dagger}U({\bf k})\hat{{\bf c}}_{{\bf k}}$,
where $U({\bf k})=e^{-iH_{2}({\bf k})T_{2}}e^{-iH_{1}({\bf k})T_{1}}$.
In Ref.~\cite{FloBigTN7}, it was found that the $U({\bf k})$ possesses
quasienergy bands with large Chern numbers and rich topological phase
diagrams due to the Floquet driving, even though the energy bands
of underlying static Hamiltonians $H_{1}({\bf k})$ and $H_{2}({\bf k})$
only carry relatively small Chern numbers. The experimental simulation
of $U({\bf k})$ and the detection of its Floquet-band Chern numbers
were also conducted recently in a setup containing a nitrogen-vacancy
center in diamond \cite{LargeCNExp}.

\begin{figure}
	\begin{centering}
		\includegraphics[scale=0.49]{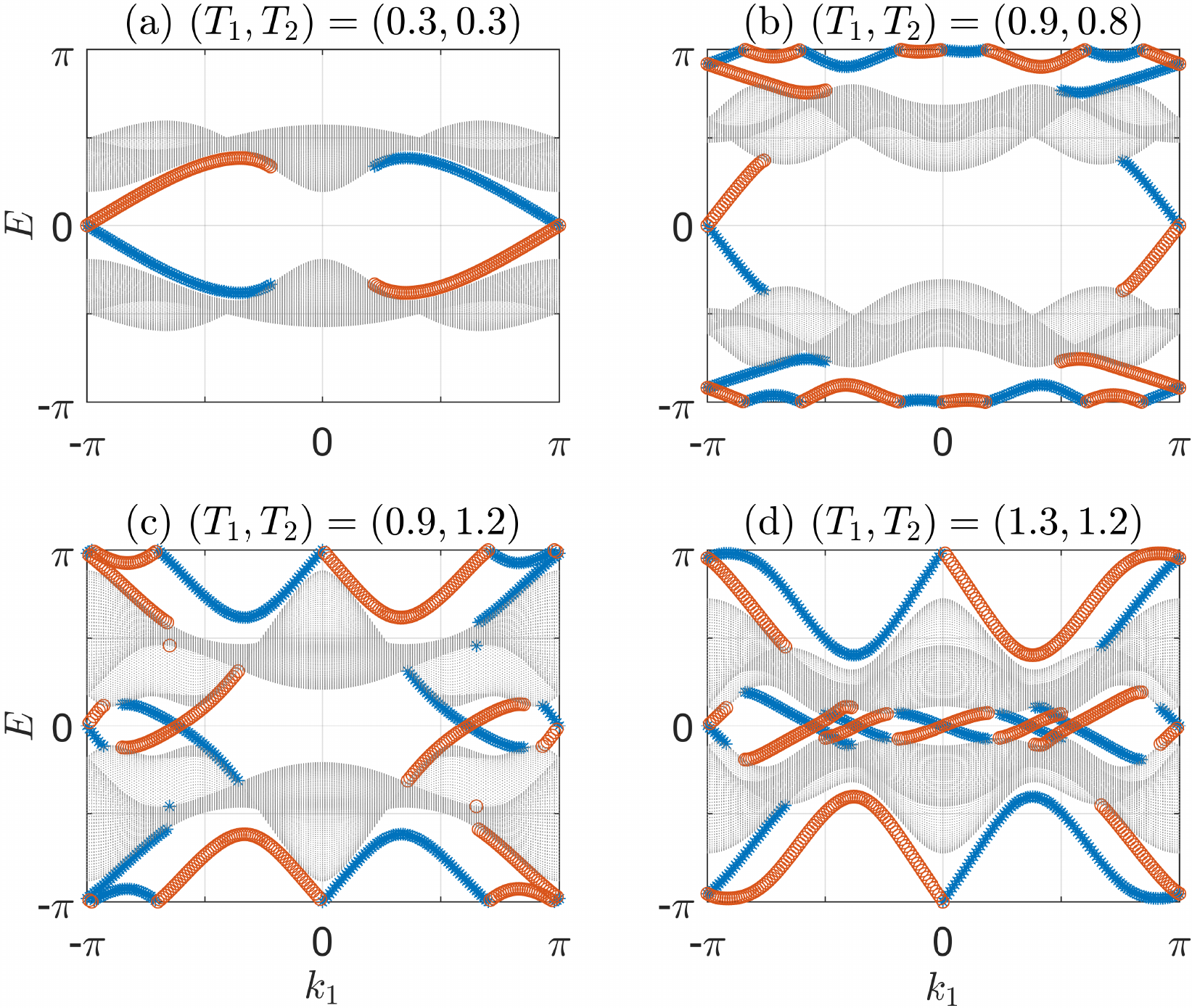}
		\par\end{centering}
	\caption{Quasienergy spectrum of the PQGHM versus the quasimomentum $k_{1}$. The
		PBC (OBC) is taken along the direction $1$ ($2$) of the lattice.
		Other system parameters are set as $t_{1}=1$, $t_{2}=0.8$, $t_{31}=0.75$,
		$t_{32}=-0.75$, $\phi_{1}=-\pi/6$, $\phi_{2}=-\pi/2$, and the length
		of lattice along the direction $2$ is $L=200$ for all panels. Blue
		stars (red circles) denote states localized around the left (right)
		edge of the lattice. \label{fig:GHM-E}}
\end{figure}

Performing a Fourier transformation $\hat{{\bf c}}_{{\bf k}}=\frac{1}{\sqrt{L_{2}}}\sum_{n=1}^{L_2}e^{-ik_{2}n}\hat{{\bf c}}_{k_{1},n}$
from momentum to position representation along the $k_{2}$-direction,
we obtain the Hamiltonian $\hat{H}=\sum_{k_{1}}(\hat{h}_{k_{1}}^{x}\sigma_{x}+\hat{h}_{k_{1}}^{y}\sigma_{y}+\hat{h}_{k_{1}}^{z}\sigma_{z})$,
where
\begin{alignat}{1}
	\hat{h}_{k_{1}}^{x}= & \sum_{n}t_{1}\left[(1+\cos k_{1})\hat{{\bf c}}_{k_{1},n}^{\dagger}\hat{{\bf c}}_{k_{1},n}+\frac{\hat{{\bf c}}_{k_{1},n}^{\dagger}\hat{{\bf c}}_{k_{1},n+1}+{\rm H.c.}}{2}\right]\nonumber \\
	+ & \sum_{n}\frac{3t_{3}}{2}\cos k_{1}\left(\hat{{\bf c}}_{k_{1},n}^{\dagger}\hat{{\bf c}}_{k_{1},n+1}+{\rm H.c.}\right)\nonumber \\
	+ & \sum_{n}\frac{t_{3}}{2i}\sin k_{1}\left(\hat{{\bf c}}_{k_{1},n}^{\dagger}\hat{{\bf c}}_{k_{1},n+1}-{\rm H.c.}\right),\label{eq:hk1x}
\end{alignat}

\begin{alignat}{1}
	\hat{h}_{k_{1}}^{y}= & \sum_{n}t_{1}\left[\sin k_{1}\hat{{\bf c}}_{k_{1},n}^{\dagger}\hat{{\bf c}}_{k_{1},n}-\frac{i}{2}\left(\hat{{\bf c}}_{k_{1},n}^{\dagger}\hat{{\bf c}}_{k_{1},n+1}-{\rm H.c.}\right)\right]\nonumber \\
	+ & \sum_{n}\frac{t_{3}}{2}\sin k_{1}\left(\hat{{\bf c}}_{k_{1},n}^{\dagger}\hat{{\bf c}}_{k_{1},n+1}+{\rm H.c.}\right)\nonumber \\
	+ & \sum_{n}\frac{t_{3}}{2i}\cos k_{1}\left(\hat{{\bf c}}_{k_{1},n}^{\dagger}\hat{{\bf c}}_{k_{1},n+1}-{\rm H.c.}\right),\label{eq:hk1y}
\end{alignat}

\begin{alignat}{1}
	\hat{h}_{k_{1}}^{z}= & \sum_{n}t_{2}\sin\phi\left[2\sin k_{1}\hat{{\bf c}}_{k_{1},n}^{\dagger}\hat{{\bf c}}_{k_{1},n}+i(\hat{{\bf c}}_{k_{1},n}^{\dagger}\hat{{\bf c}}_{k_{1},n+1}-{\rm H.c.})\right]\nonumber \\
	- & \sum_{n}t_{2}\sin\phi\sin k_{1}\left(\hat{{\bf c}}_{k_{1},n}^{\dagger}\hat{{\bf c}}_{k_{1},n+1}+{\rm H.c.}\right)\nonumber \\
	- & \sum_{n}it_{2}\sin\phi\cos k_{1}\left(\hat{{\bf c}}_{k_{1},n}^{\dagger}\hat{{\bf c}}_{k_{1},n+1}-{\rm H.c.}\right).\label{eq:hk1z}
\end{alignat}

Following the above mentioned quench protocol, we obtain the quasienergy
spectrum of Floquet operator $\hat{U}=e^{-i\hat{H}_{2}T_{2}}e^{-i\hat{H}_{1}T_{1}}$
under the PBC (OBC) along the direction $1$ ($2$) of the lattice, as
shown in Fig.~\ref{fig:GHM-E}. For all considered cases, we find
two Floquet bulk bands connected by gap-traversing chiral edge states.
In the first quasienergy Brillouin zone, the lower and upper bands
have Chern numbers $\pm1$, $\pm2$, $\pm4$ and $\pm7$ in 
Figs.~\ref{fig:GHM-E}(a), (b), (c) and (d), which are equal to the net
difference between the numbers of chiral edge states above and below
a band while residing at the same edge. This is consistent with the
bulk-edge correspondence of 2D Floquet Chern insulators revealed at
the spectral level \cite{FloBigTN7}. The emerging Floquet bands with
large Chern numbers $\pm4$ and $\pm7$ are unreachable in the static
limit of the model \cite{LargeCNExp}. It reflects one key advantage
of Floquet engineering, i.e., to create phases with large topological
invariants and many edge states by simply tuning the time durations
of the drive.

\begin{figure}
	\begin{centering}
		\includegraphics[scale=0.49]{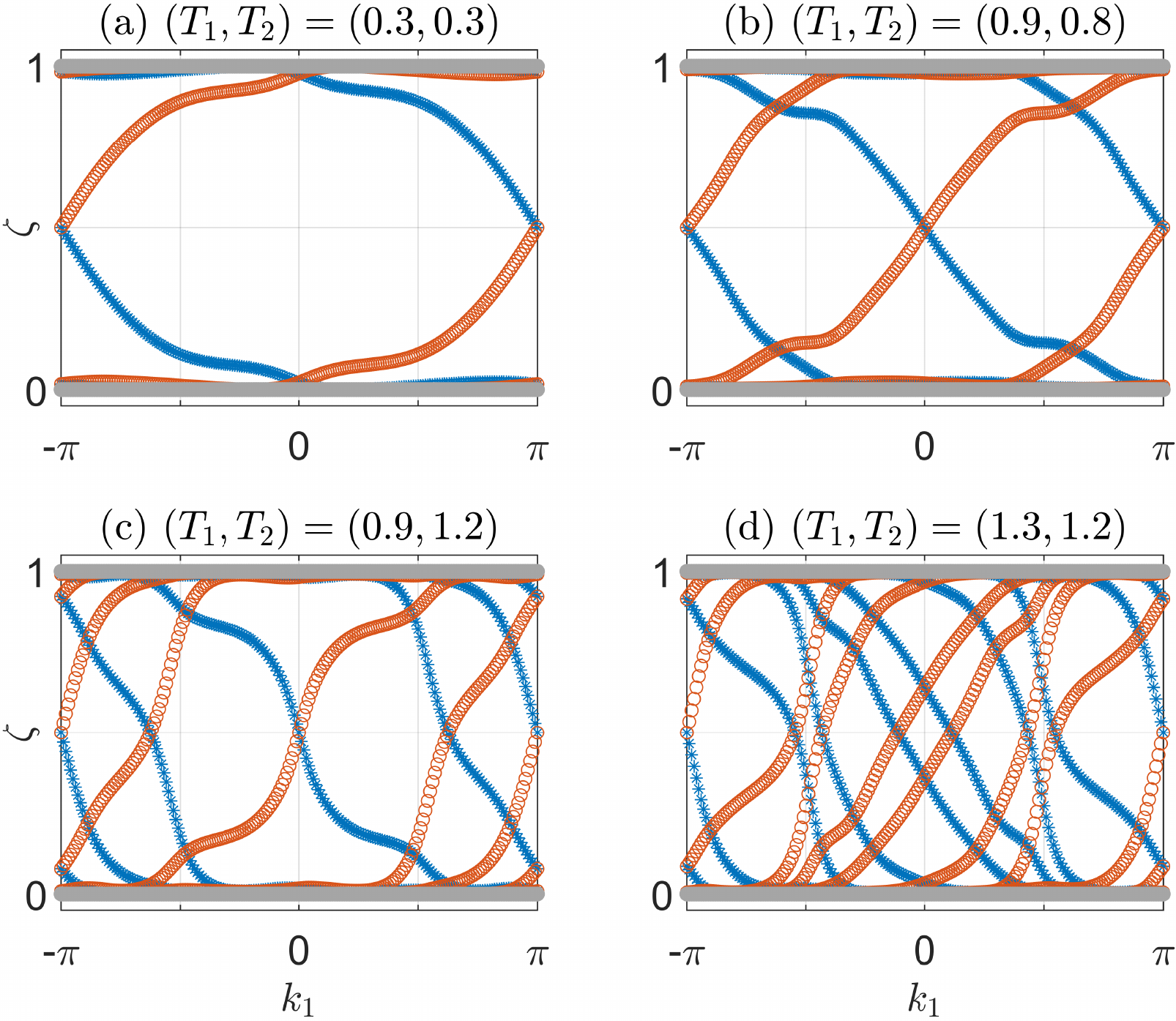}
		\par\end{centering}
	\caption{Correlation matrix spectra of the PQGHM versus the quasimomentum
		$k_{1}$. PBCs are taken along both directions of
		the lattice. Other system parameters and the length of lattice are
		chosen to be the same as in Fig.~\ref{fig:GHM-E}. Blue stars (red
		circles) denote eigenmodes localized around the left (right) entanglement
		cuts between the subsystems A and B. \label{fig:GHM-ES}}
\end{figure}

We now uncover the topology, phase transitions and bulk-edge correspondence
of the PQGHM from its ES, EE and the related topological numbers.
We first take the PBC along direction $2$ by setting $\hat{{\bf c}}_{k_{1},n+L_{2}}=\hat{{\bf c}}_{k_{1},n}$,
where $L_{2}$ is the number of unit cells along the second dimension
of the lattice. Then we decompose the system into two equal segments
A and B, and obtain the ES of reduced density matrix $\hat{\rho}_{{\rm A}}$
following the Eqs.~(\ref{eq:GS})\textendash (\ref{eq:ES}) in 
Sec.~\ref{sec:The}. Here the state $|\Psi\rangle$ in Eq.~(\ref{eq:GS})
describes the uniformly filled lower Floquet band in the first quasienergy
Brillouin zone. The resulting ES is presented in Figs.~\ref{fig:GHM-ES}(a)\textendash (d),
with the same system parameters as those used in Figs.~\ref{fig:GHM-E}(a)\textendash (d).
We observe that large amounts of the correlation matrix eigenvalues
$\zeta_{j}$ {[}Eq.~(\ref{eq:CM}){]} reside near $\zeta=0$ and $\zeta=1$,
making vanishing contributions to the EE in Eq.~(\ref{eq:EE}). In
the meantime, chiral ES flows across the ES gap are observed in each
panel of Fig.~\ref{fig:GHM-ES}. These chiral bands are formed by
eigenstates of the Floquet entanglement Hamiltonian $\hat{H}_{{\rm A}}$
{[}Eq.~(\ref{eq:HA}){]} that are localized around the entanglement
cuts between subsystems A and B. Furthermore, the number of chiral
edge bands $n_{{\rm c}}$ at the same edge in the ES is found to be
equal to the Chern number ${\cal C}_{0}$ of the Floquet band used
the in the construction of the initial density matrix $\hat{\rho}$
[Eq.~(\ref{eq:GS})], i.e., 
\begin{equation}
	n_{{\rm c}}=|{\cal C}_{0}|.\label{eq:BBC0}
\end{equation}
This relation establishes the correspondence between the Chern topology
of Floquet bands and the number of chiral edge modes in the ES for
2D Floquet Chern insulators with two bands. As demonstrated in 
Sec.~\ref{subsec:KHM}, it is also valid for systems with multiple bands.

\begin{figure}
	\begin{centering}
		\includegraphics[scale=0.49]{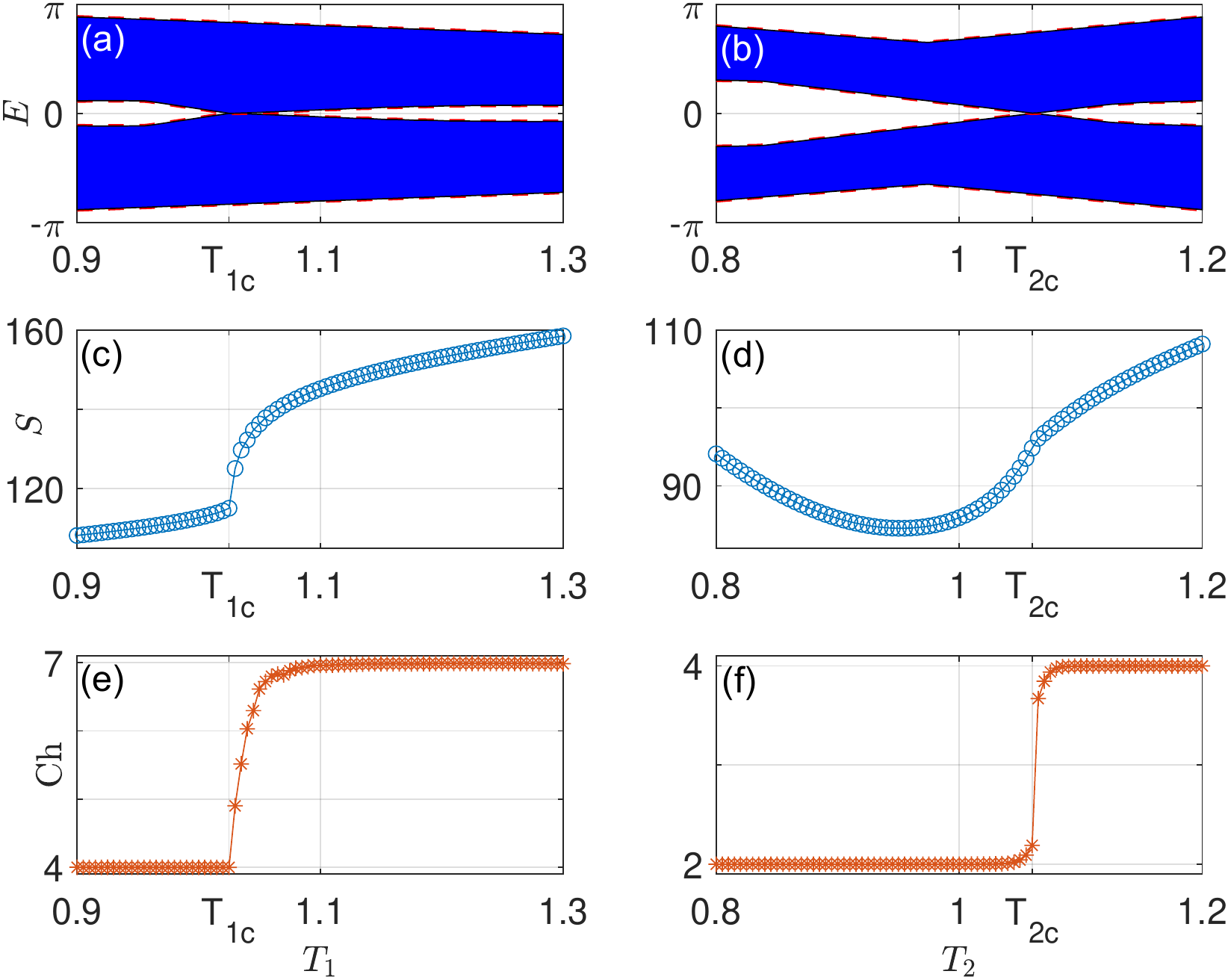}
		\par\end{centering}
	\caption{Quasienergy spectra {[}in (a), (b){]}, EE {[}in (c), (d){]} and real-space
		Chern numbers {[}in (e), (f){]} of the PQGHM versus the driving time
		durations $T_{1}$ and $T_{2}$. The size of lattice is $L_{1}=L_{2}=60$.
		PBCs are taken along both directions of the lattice. We set $T_{2}=1.2$
		for (a), (c), (e) and $T_{1}=0.9$ for (b), (d), (f). Other system
		parameters are chosen to be the same as those used in Fig.~\ref{fig:GHM-E}.
		\label{fig:GHM-EE-CN}}
\end{figure}

We next consider topological phase transitions in the system. In 
Figs.~\ref{fig:GHM-EE-CN}(a) and \ref{fig:GHM-EE-CN}(b), we show the Floquet
spectra under the PBC versus the quench time spans $T_{1}$ and $T_{2}$.
The gaps between the two quasienergy bands close when $T_{1}=T_{1{\rm c}}$
and $T_{2}=T_{2{\rm c}}$ in Figs.~\ref{fig:GHM-EE-CN}(a) and \ref{fig:GHM-EE-CN}(b),
respectively. The critical values $T_{1{\rm c}}$ and $T_{2{\rm c}}$
can be obtained by solving the equations of gapless condition, i.e.,
${\bf h}_{1}/|{\bf h}_{1}|={\bf h}_{2}/|{\bf h}_{2}|$ and $T_{1}|{\bf h}_{1}|+T_{2}|{\bf h}_{2}|=\ell\pi$
($\ell\in\mathbb{Z}$). Here ${\bf h}_{1}$ and ${\bf h}_{2}$ are
${\bf k}$-dependent vectors formed by the three components in front
of the Pauli matrices in $H_{1}({\bf k})$ and $H_{2}({\bf k})$,
respectively \cite{LargeCNExp}. With the increase of $T_{1}$ ($T_{2}$)
from $T_{1}-0^{+}$ ($T_{2}-0^{+}$) to $T_{1}+0^{+}$ ($T_{2}+0^{+}$),
the PQGHM undergoes a topological phase transition accompanied by
the quantized change of Floquet band Chern numbers from $\pm4$ ($\pm2$)
to $\pm7$ ($\pm4$) \cite{LargeCNExp}. The EE curves in Figs.~\ref{fig:GHM-EE-CN}(c)
and \ref{fig:GHM-EE-CN}(d) are obtained following Eqs.~(\ref{eq:CM})\textendash (\ref{eq:EE}).
We find that the derivatives $\partial_{T_{1}}S$ and $\partial_{T_{2}}S$
become discontinuous at $T_{1}=T_{1{\rm c}}$ and $T_{2}=T_{2{\rm c}}$,
signifying the transitions between different Floquet Chern insulator
phases from the perspective of EE.

Finally, we characterize the topology of Floquet entanglement Hamiltonian
$H_{{\rm A}}$ {[}Eq.~(\ref{eq:HA}){]} and relate it to the edge
states observed in the quasienergy spectrum. To this end, we consider
the real-space Chern number Ch defined through Eqs.~(\ref{eq:LCM})\textendash (\ref{eq:CN}).
In Figs.~\ref{fig:GHM-EE-CN}(e) and \ref{fig:GHM-EE-CN}(f), we show
the Ch thus computed at different values of quenching time intervals
$T_{1}$ and $T_{2}$ for the PQGHM. Quantized jumps are observed
in the Ch around $T_{1}=T_{1{\rm c}}$ and $T_{2}=T_{2{\rm c}}$,
i.e., at the expected transition points between different Floquet
Chern insulator phases. Away from the transition points, the real-space
Chern numbers are found to take quantized values, which are further
equal to the Chern numbers ${\cal C}_0$ of the occupied Floquet bands employed
in the construction of reduced density matrix $\hat{\rho}_{{\rm A}}$.
Therefore, we arrive at the relationship between the Chern numbers
of entanglement Hamiltonian and the chiral edge modes in the quasienergy
spectrum, i.e., 
\begin{equation}
	|n_{{\rm L}}-n'_{{\rm L}}|=|{\rm Ch}|,\label{eq:BBC3}
\end{equation}
where $n_{{\rm L}}$ and $n'_{{\rm L}}$ denote the net number of
chiral edge bands leaving and entering the occupied Floquet band at
the left edge in the lattice (see Fig.~\ref{fig:GHM-E}). We refer
to this relation as an entanglement bulk-edge correspondence of 2D
Floquet Chern insulators. As one key result of this work, it establishes
the connection between the Chern topology of Floquet entanglement
Hamiltonian under the PBC and Floquet chiral edge states under the
OBC. All the topological information of a Floquet Chern band can thus be
extracted from the entanglement nature of its related reduced density matrix.
Eq.~(\ref{eq:BBC3}) can be further generalized to Floquet Chern
insulators with multiple quasienergy bands, as shown in the following
subsection.

\subsubsection{Kicked quantum Hall insulator\label{subsec:KHM}}

We finally consider fermions hopping in a square lattice with perpendicular
magnetic fluxes. The system Hamiltonian is made time-dependent by
adding periodic delta kicks to the hopping amplitudes along $y$-direction
of the lattice, yielding
\begin{alignat}{1}
	\hat{H}(t)= & \frac{J}{2}\sum_{x,y}\left(\hat{c}_{x+1,y}^{\dagger}\hat{c}_{x,y}+{\rm H.c.}\right)\nonumber \\
	+ & \frac{V}{2}\sum_{x,y}\left(e^{i\lambda x}\hat{c}_{x,y+1}^{\dagger}\hat{c}_{x,y}+{\rm H.c.}\right)\sum_{\ell\in\mathbb{Z}}\delta(t-\ell).\label{eq:KHMH}
\end{alignat}
Here $\hat{c}_{x,y}$ ($\hat{c}_{x,y}^{\dagger}$) annihilates (creates)
a fermion on the lattice site $(x,y)$. $J$ and $V$ are hopping
amplitudes along $x$ and $y$ directions. $\lambda$ determines the
magnetic flux over each cell of the lattice. The driving period is
set to $T=1$. The Floquet operator of the system then takes the form
\begin{alignat}{1}
	\hat{U}& = e^{-\frac{iV}{2}\sum_{x,y}\left(e^{i\lambda x}\hat{c}_{x,y+1}^{\dagger}\hat{c}_{x,y}+{\rm H.c.}\right)}\nonumber \\
	& \times e^{-\frac{iJ}{2}\sum_{x,y}\left(\hat{c}_{x+1,y}^{\dagger}\hat{c}_{x,y}+{\rm H.c.}\right)}.\label{eq:KHMU}
\end{alignat}
It is usually called the kicked Harper model (KHM) or kicked quantum
Hall system \cite{FloBigTN3,FloBigTN4,FloBigTN5}. For $\lambda=2\pi p/q$,
with $p$ and $q$ being coprime integers, the Floquet spectrum of
$\hat{U}$ groups into $q$ quasienergy bands under the PBC. In this
work, we focus on the case with $p/q=1/3$, so that there are three
Floquet bands in the quasienergy spectrum. The topology of each Floquet
band is characterized by an integer-quantized Chern number \cite{FloBigTN3}.
With the change of $J$ and/or $V$, the KHM can undergo topological
phase transitions between different Floquet Chern insulator phases,
which are accompanied by quantized changes of band Chern numbers \cite{FloBigTN3}.
Floquet bands with large Chern numbers were found at large hopping
amplitudes $J$ and/or $V$ \cite{FloBigTN3}, showing the advantage
of Floquet engineering in creating phases with strong topological
signals.

\begin{figure}
	\begin{centering}
		\includegraphics[scale=0.49]{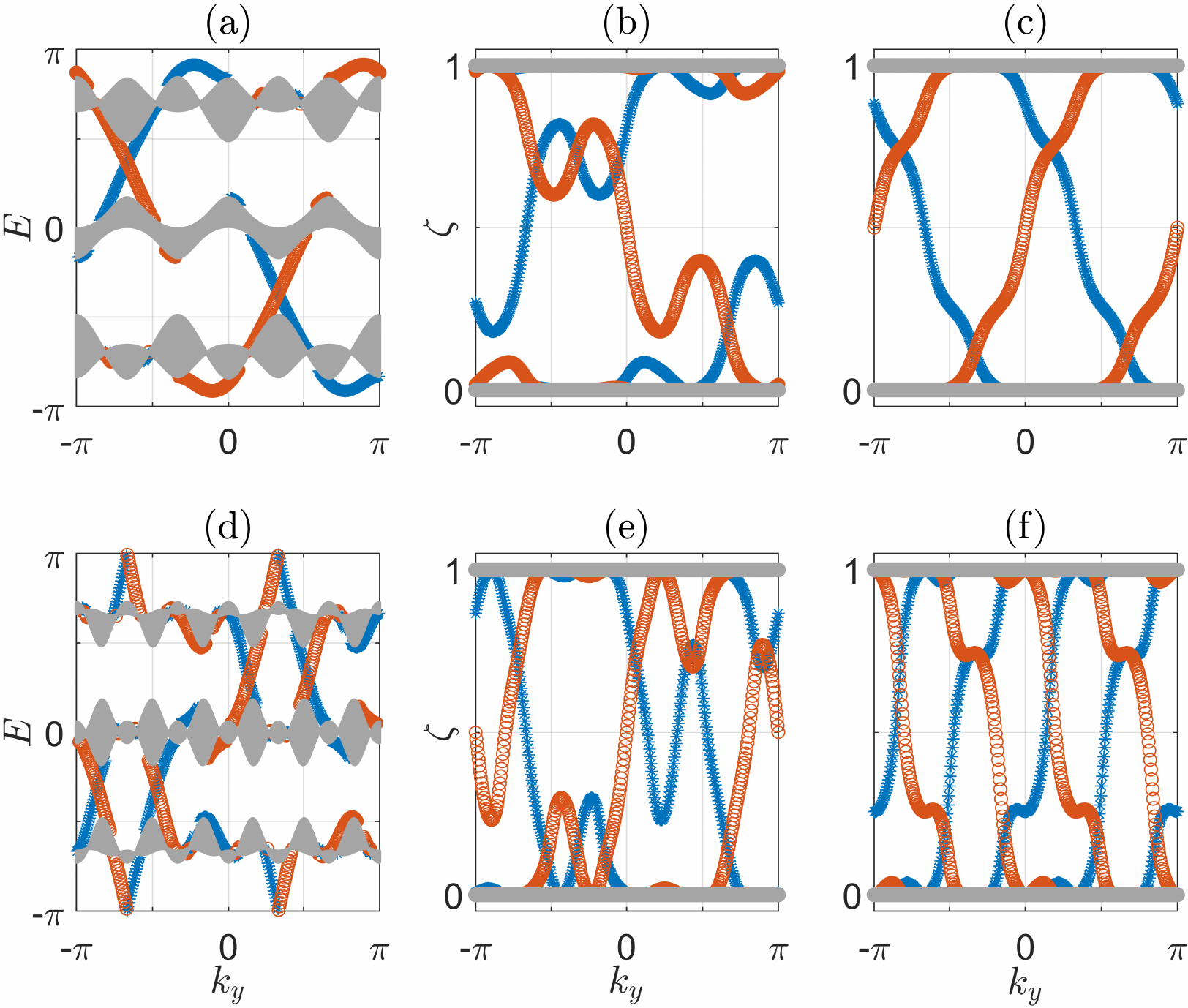}
		\par\end{centering}
	\caption{Quasienergy spectrum $E$ and correlation matrix spectrum $\zeta$
		of the KHM versus the quasimomentum $k_{y}$. The length of lattice
		is chosen to be $L=300$ along the $x$-direction for all panels.
		System parameters are set as $(J,V)=(2\pi/3,\pi)$ for (a)\textendash (c)
		and $(J,V)=(2\pi/3,2\pi)$ for (d)\textendash (f). The OBC (PBC) is
		taken along the $x$-direction in the calculation of $E$ ($\zeta$).
		Blue stars (red circles) denote states localized around the left (right)
		edges in (a), (d), and states localized around the left (right) entanglement
		cuts between the subsystems A and B in (b), (c), (e), (f), respectively.
		For (b) and (e), the states $|\Psi\rangle$ correspond to the uniform
		filling of the bottom Floquet bands in (a) and (d). For (c) and (f),
		the states $|\Psi\rangle$ correspond to the uniform filling of the
		middle Floquet bands in (a) and (d). \label{fig:KHM-E-ES}}
\end{figure}

Taking the OBC (PBC) along the $x$ ($y$) direction and performing the
Fourier transformation $\hat{c}_{x,y}=\frac{1}{\sqrt{L_{y}}}\sum_{k_{y}}e^{ik_{y}y}\hat{c}_{x,k_{y}}$
along $y$ direction, we can express the Floquet operator of the KHM
as $\hat{U}=\sum_{k_{y}}\hat{U}(k_{y})$, where
\begin{alignat}{1}
	\hat{U}(k_{y})& = e^{-iV\sum_{x}\cos(\lambda x-k_{y})\hat{c}_{x,k_{y}}^{\dagger}\hat{c}_{x,k_{y}}}\nonumber \\
	& \times  e^{-\frac{iJ}{2}\sum_{x}\left(\hat{c}_{x+1,k_{y}}^{\dagger}\hat{c}_{x,k_{y}}+{\rm H.c.}\right)}.\label{eq:KHMUy}
\end{alignat}
$k_{y}\in[-\pi,\pi)$ is the quasimomentum along $y$. Typical examples
of the Floquet spectrum of $\hat{U}(k_{y})$ are shown in Figs.~\ref{fig:KHM-E-ES}(a)
and \ref{fig:KHM-E-ES}(d). The Chern numbers of the three Floquet
bands from bottom to above in Figs.~\ref{fig:KHM-E-ES}(a) and \ref{fig:KHM-E-ES}(d)
are $(1,-2,1)$ and $(-2,4,-2)$ \cite{FloBigTN3}. This is coincide
with the difference between the number of chiral edge bands at the
same edge and with the same chirality leaving and entering a given
Floquet band, which verifies the bulk-edge correspondence of Floquet
Chern insulators from the aspect of quasienergy spectrum \cite{FloBigTN3}.

We now unveil the topological nature and bulk-edge correspondence
of the KHM by investigating its ES and EE. Starting with the Floquet
operator $\hat{U}(k_{y})$ in Eq.~(\ref{eq:KHMUy}), we take the PBC
along $x$-direction and let the density matrix in Eq.~(\ref{eq:R})
to describe the many-particle Floquet state that fills the bottom
or the middle Floquet band of $\hat{U}(k_{y})$ uniformly. Bisecting
the system into two equal parts and following the steps in Eqs.~(\ref{eq:RA})\textendash (\ref{eq:ES}),
we obtain the ES of KHM as shown in Figs.~\ref{fig:KHM-E-ES}(b),
\ref{fig:KHM-E-ES}(c), \ref{fig:KHM-E-ES}(e), and \ref{fig:KHM-E-ES}(f).
In each case, we find that most of the correlation matrix eigenvalues
are pinned around $\zeta=0$ and $\zeta=1$, yielding vanishing contributions
to the EE in Eq.~(\ref{eq:EE}). Meanwhile, eigenmodes traversing
the gap of ES are found and they are localized at the entanglement
cuts between the subsystems A and B. Moreover, the ES of these eigenmodes
are chiral, and the number of such chiral entanglement bands at each
edge is equal to the Chern number of the filled Floquet band, which
verifies the Eq.~(\ref{eq:BBC0}). This observation allows us to inspect
the Chern topology of a Floquet band and the related edge states from
its ES. It also demonstrates that our approach to the ES of Floquet
Chern bands are not limited to two-band models, but works for Floquet
Chern insulators with any number of gapped quasienergy bands.

\begin{figure}
	\begin{centering}
		\includegraphics[scale=0.49]{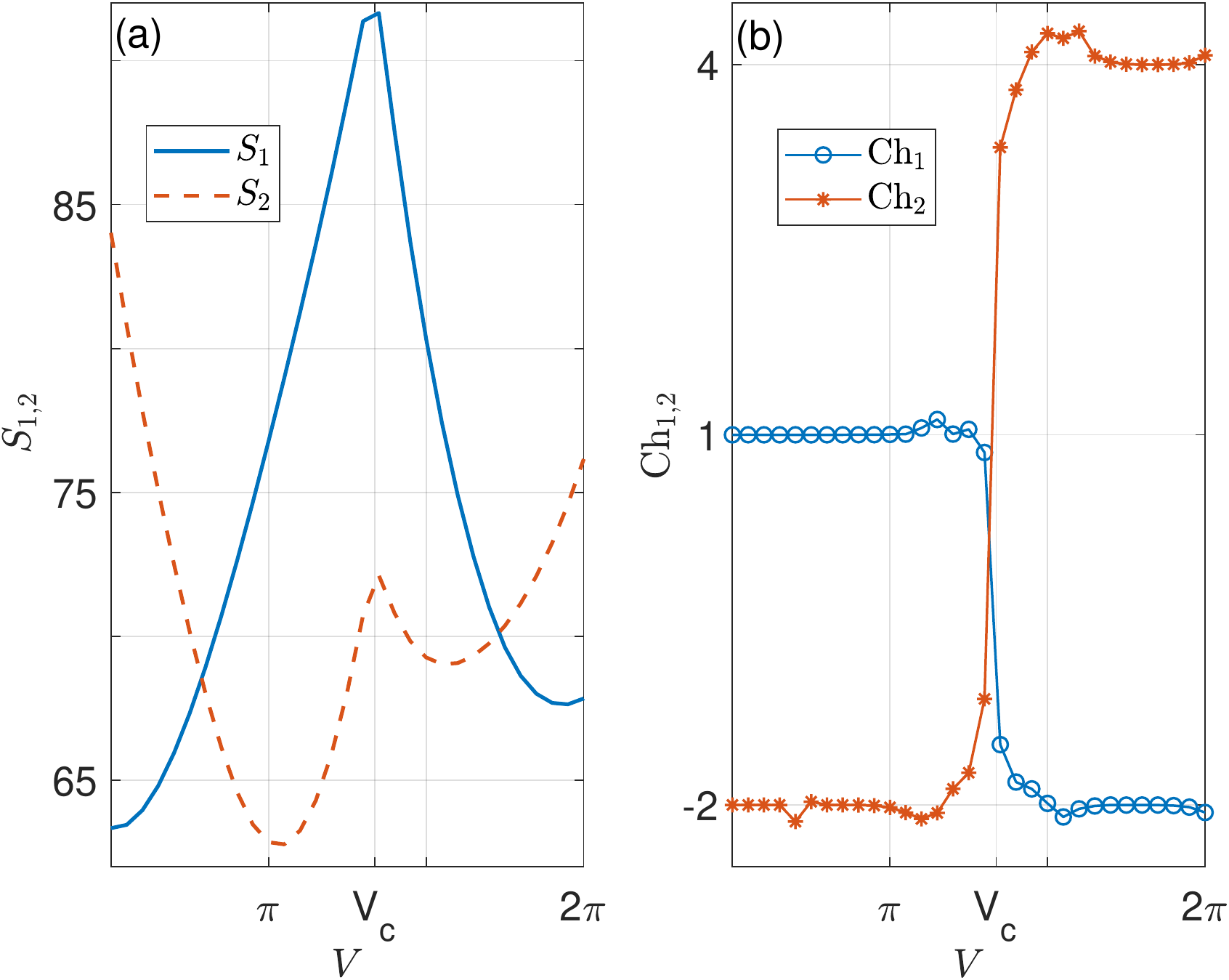}
		\par\end{centering}
	\caption{The EE and real-space Chern numbers of the entanglement Hamiltonian for
		the KHM. The hopping amplitude along $x$ direction is set to $J=2\pi/3$.
		The size of lattice is $L_{x}=L_{y}=72$ with PBCs taken along both
		$x$ and $y$ directions. In (a) and (b), $S_{1}$ ($S_{2}$) and
		${\rm Ch}_{1}$ (${\rm Ch}_{2}$) refer to the EE and the real-space
		Chern number of the entanglement Hamiltonian ${\hat H}_{{\rm A}}$ when the
		bottom (middle) quasienergy band is filled in the construction of
		$\hat{\rho}_{{\rm A}}$. \label{fig:KHM-EE-CN}}
\end{figure}

The difference between the band Chern numbers in Figs.~\ref{fig:KHM-E-ES}(a)
and \ref{fig:KHM-E-ES}(d) suggests that there is at least a transition
from one Floquet Chern insulator phase to another when the hopping
amplitude $V$ goes from $\pi$ to $2\pi$. In Fig.~\ref{fig:KHM-EE-CN}(a),
we show the EE of the bottom and middle filled Floquet band of the KHM
versus $V$ under the PBC along both two spatial dimensions 
{[}Eq.~(\ref{eq:EE}){]}. We indeed observe nonanalytic cusps in $S_{1}$
and $S_{2}$ near $V=V_{\rm c}$, which is coincident with the topological
phase transition point of the KHM \cite{WangPRE2013}. To characterize
the topology of this transition and the bulk-edge correspondence directly
at the level of Floquet entanglement Hamiltonian {[}Eq.~(\ref{eq:HA}){]},
we compute the real-space Chern number of $\hat{H}_{{\rm A}}$ at
half-filling under the PBC from the local Chern marker following 
Eqs.~(\ref{eq:LCM})\textendash (\ref{eq:CN}). The results for the cases
of filled bottom and middle Floquet bands of the KHM are displayed
by ${\rm Ch}_{1}$ (stars) and ${\rm Ch}_{2}$ (circles) in Fig.~\ref{fig:KHM-EE-CN}(b).
We find that both the Chern numbers show quantized jumps at the transition
point $V=V_{\rm c}$, implying that they can correctly capture the topological
phase transition in the system. Moreover, away from the transition
point, ${\rm Ch}_{1}$ and ${\rm Ch}_{2}$ take quantized values (up
to fluctuations due to finite-size effects) that are equal to the
Chern numbers of the bottom and middle Floquet bands in Figs.~\ref{fig:KHM-E-ES}(a)
and \ref{fig:KHM-E-ES}(d). Therefore, we arrive at the following
entanglement bulk-edge correspondence for Floquet Chern insulators
\begin{equation}
	|n_{{\rm L}}-n'_{{\rm L}}|=|{\rm Ch}_{j}|.\label{eq:BBC4}
\end{equation}
Here $n_{{\rm L}}$ and $n'_{{\rm L}}$ represent the net number of
chiral edge bands leaving and entering the $j$th Floquet quasienergy
band at the left edge of the lattice. ${\rm Ch}_{j}$ denotes the
real-space Chern number of the Floquet entanglement Hamiltonian, which
is originated from the reduced density operator of the $j$th Floquet band.
Eq.~(\ref{eq:BBC4}) generalizes the Eq.~(\ref{eq:BBC3}) to Floquet
Chern insulators with more than two bands. We have explored other
parameter regions of the KHM and obtain results that are consistent
with those reported in Figs.~\ref{fig:KHM-E-ES}, \ref{fig:KHM-EE-CN}
and by Eq.~(\ref{eq:BBC4}). Therefore, we conclude that the entanglement
tools introduced in this work could characterize the topology, edge
states and bulk-edge correspondence of 2D Floquet Chern insulators
with an arbitrary number of quasienergy bands.

\section{Discussion and Conclusion\label{sec:Sum}}

It is interesting to discuss the connections and differences between our work and some previous studies.
First, in 1D Floquet topological insulators, both the zero and $\pi$ quasienergy edge modes show their signatures in the ES. The $\pi$ modes do not exist in usual static topological systems. Their entanglement signatures are thus unique to Floquet systems showcased by the first and second models in Sec.~\ref{sec:Res}.
Second, in some former studies \cite{ESEETP52,ESEETP6,ESEETP7,ESEETP8,ESEETP9,ESEETP10}, the ES of usual Chern and quantum Hall insulators were investigated. The connection between the chiral edge states in the ES and that in the energy spectrum was demonstrated. Topological invariants were also introduced to characterize the ES in momentum space. In our work, we considered the ES and EE of periodically driven Chern and quantum Hall insulators, whose topological properties are carried by the underlying Floquet states instead of a ground state in usual static topological systems. Even though the chiral edge states observed in the Floquet ES are similar to those found in static topological models, we have introduced topological invariants to characterize the bulk-edge correspondence between these edge modes and the topology of Floquet entanglement Hamiltonian directly in real space, which is different from previous approaches.
Third, non-analytical properties of EE across quantum phase transitions have been identified in early studies \cite{EntangleRev1}. The behaviors of EE around the transition points between different topological phases were also explored in usual static systems \cite{EETPT1,EETPT2,EETPT3,EETPT4}. There, non-analytic signatures in the EE were observed and the scaling properties of EE were investigated. In our study, we also observed non-analytic signatures in the EE of Floquet states when the underlying system goes from one Floquet topological insulator phase to another. Compared with previous results, we tend to believe that in general, the EE possesses similar features around a transition point between two topological insulator/superconductor phases either in usual static systems or in Floquet driven systems. In our case studies, we focus more on the topological characterization of ES and its related bulk-edge correspondence in Floquet systems. The EE are presented mainly to provide complementary information for the signatures of Floquet topological phase transitions. We have supplemented our results with further details about EE in the Appendix \ref{app:A}.

In Ref.~\cite{YatesPRB2016}, the ES and EE of a graphene lattice under different Floquet driving protocols were explored with a theoretical formalism tailored to two-band models and a focus on the time-dependent properties of the entanglement measures. Signatures of chiral edge states in the ES were also identified there. In our work, we instead focus on the ES and EE of filled Floquet bands and present a framework that is applicable to study them in noninteracting fermionic systems with more than two quasienergy bands and large Chern numbers, as illustrated by the case studies in Sec.~\ref{subsec:ESEE2D}. Moreover, we characterized the changes of ES, EE and topological property of entanglement Hamiltonian when the system undergoes transitions between different Floquet Chern or quantum Hall insulator phases, which were not considered in Ref.~\cite{YatesPRB2016}. In Refs.~\cite{YatesPRB2017} and \cite{YatesPRL2018}, the ES, EE and central charge of a 1D two-band Floquet topological superconductor were investigated by a theory presented in terms of Majorana operators. Non-analytic behaviors in EE when the system undergoes transitions between different Floquet superconducting phases were studied. In our work, we formalized our theory in terms of normal fermions and demonstrated it in two 1D Floquet topological insulator models with different numbers of quasienergy bands and belonging to symmetry classes that are different from those explored in Refs.~\cite{YatesPRB2017} and \cite{YatesPRL2018}. We also clarified the bulk-edge correspondence by comparing the ES and the winding numbers of Floquet entanglement Hamiltonians in different time frames. Furthermore, the real-space Chern and winding numbers employed in our work allows us to directly characterize the topology, phase transitions and bulk-edge correspondence residing in the bulk Floquet ES, which were not considered in Refs.~\cite{YatesPRB2016}--\cite{YatesPRL2018}. Besides helping us to build the connection between the topological properties of quasienergy and entanglement spectra, these invariants are also robust to the change of boundary conditions and the presence of symmetry-preserving impurities, making them generalizable to the characterization of Floquet entanglement topology in more complicated situations.

In summary, we propose a framework to describe the ES and EE of
noninteracting fermions in Floquet systems. The theory is applicable
to Floquet lattice models with an arbitrary number of quasienergy bands and
at arbitrary fillings for each band. Open-bulk winding numbers and
real-space Chern numbers are further constructed to characterize the
topology of Floquet entanglement Hamiltonians directly for 1D and
2D systems. In one dimension, the correspondences between the topology
of bulk ES and the Floquet edge states are established for Floquet topological
insulators in the symmetry classes BDI and CII. In two dimensions,
the total Chern number of filled Floquet bands determines the number of chiral
edge modes traversing the gap of ES with the same chirality, which
is also consistent with the real-space Chern number of the Floquet
entanglement Hamiltonian. The EE further shows non-analytic behaviors
around the transition points between different Floquet topological
phases. Our work thus revealed the topological phase transition and bulk-edge correspondence of Floquet
topological matter from quantum entanglement perspectives, and
provided efficient means to characterize the topological nature of
Floquet entanglement Hamiltonians (or reduced density matrices of
Floquet states). It further demonstrated the generality and usefulness
of quantum information measures in characterizing Floquet topological
phases.

In future work, the extensions of our entanglement framework
to Floquet systems with disorder, in high spatial dimensions, with other symmetry
constraints and in gapless systems (e.g., the Floquet Weyl semimetal \cite{WSM1,WSM2})
deserve to be investigated. 
For example, beyond the on-resonance condition, the kicked rotor considered in Sec.~\ref{subsec:KRS}
does not possess the translational symmetry in momentum space, and its topological properties
may have distinct and deeper physical origins \cite{KRSOther1,KRSOther2}, such as the emergence of integer quantum Hall effect from chaos \cite{KRSTian1,KRSTian2}.
Meanwhile, the application of our entanglement approach should be relatively straightforward. One can first express the Floquet operator of the kicked rotor in a given basis and diagonalize it upon appropriate truncations. One may then pick up a set of Floquet eigenstates of the kicked rotor to form its many-particle density matrix. By decomposing the momentum space of the rotor into two parts A, B and tracing out the degrees of freedom belonging to B, one would obtain a reduced density matrix for which the ES and EE can be computed by the formulas presented in this work. Moreover, the open-bulk winding number is directly defined in the discrete lattice or momentum space, and it is applicable to systems with disorder or other types of translational symmetry breaking so long as the chiral symmetry is preserved. When the chiral symmetry is also broken, other topological invariants might be needed to characterize the Floquet entanglement Hamiltonian of the kicked rotor. Finding such an invariant should constitute an interesting future study.
As our theory is tailored to deal with
many-particle Floquet states, we can further incorporate
inter-particle interactions and decode the entanglement nature of the
emerging correlated phases in Floquet systems with potential topological features,
such as the coupled or mean-field interacting kicked rotor~\cite{INTQKR1,INTQKR2}.
For the coupled kicked rotor \cite{INTQKR2}, one may treat the coordinates of two rotors as the coordinates along two different spatial dimensions. The interaction term in the Lieb-Liniger form can then be interpreted as an onsite potential along the diagonal direction of the effective 2D lattice. We are now left with a noninteracting problem in a 2D kicked lattice. The quasienergy eigenstates of the system may then be obtained by diagonalizing the Floquet operator of the effective 2D model. Following these treatments, our entanglement approach might be implemented in parallel with what we have done for the 2D models in this work.
Finally, it deserves to mention that similar to the work of Peschel \cite{EntangleRev4}, our approach of obtaining the ES and EE from the single-particle correlation matrix is restricted to many-body states in Gaussian form. Finding a clear path to extend this approach to non-Gaussian states or other more complicated situations is a challenging yet very intriguing topic for future explorations.

\begin{acknowledgments}
We acknowledge Jiangbin Gong and Chushun Tian for helpful comments. This work is supported by the National Natural Science Foundation of China (Grants No.~12275260 and No.~11905211), the Young Talents Project at Ocean University of China (Grant No.~861801013196), and the Applied Research Project of Postdoctoral Fellows in Qingdao (Grant No.~861905040009).
\end{acknowledgments}

\appendix
\section{EE around Floquet topological phase transition points}\label{app:A}

\begin{figure}
	\begin{centering}
		\includegraphics[scale=0.48]{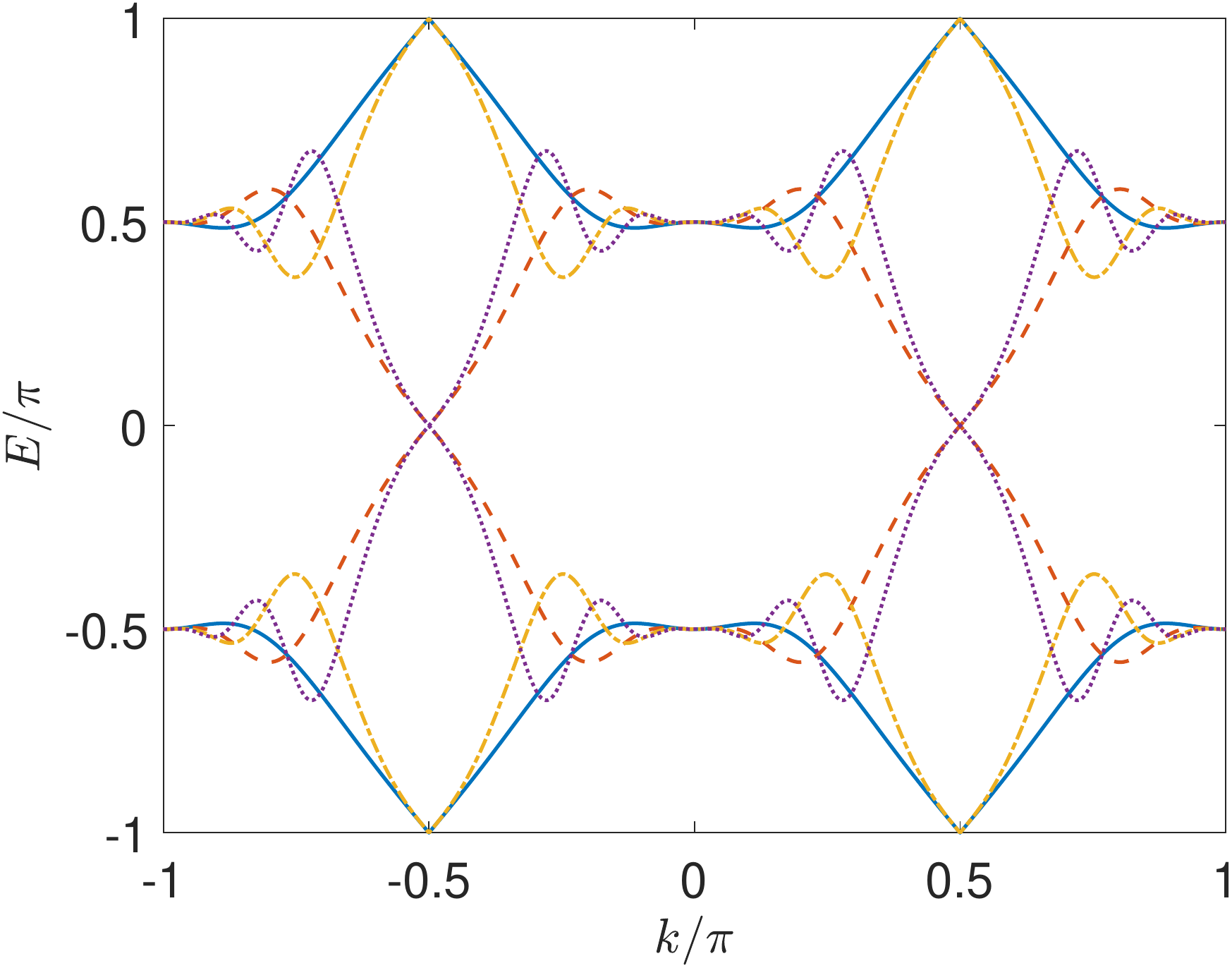}
		\par\end{centering}
	\caption{Floquet bands of the spin-$1/2$ ORDKR under the PBC. We set $K_1=0.5\pi$ for all cases. The solid, dashed, dash-dotted and dotted curves represent the quasienergy bands for $K_2=\pi$, $2\pi$, $3\pi$, and $4\pi$, respectively. In each case, the lower band is filled in the calculation of EE.}\label{fig:KRS-Evsk}
\end{figure}

In this Appendix, we elaborate a bit more on the EE of the spin-$1/2$ ORDKR in Sec.~\ref{subsec:KRS}. Taking the PBC and performing the Fourier transformations $\hat{\bf c}_n^{(\dagger)}=\frac{1}{\sqrt{L}}\sum_k e^{(-)i k n}\hat{\bf c}_k^{(\dagger)}$ with the quasiposition $k\in[-\pi,\pi)$, we find the Floquet operator Eq.~(\ref{eq:KRSU}) in $k$-space to be $U(k)=e^{-iK_2\sin k\sigma_y}e^{-iK_1\cos k\sigma_x}$ \cite{KRSZhou2018}. Solving the eigenvalue equation $U(k)|\psi\rangle=e^{-iE(k)}|\psi\rangle$, we find two Floquet quasienergy bands with the dispersions $E_\pm(k)=\pm\arccos[\cos(K_1\cos k)\cos(K_2\sin k)]$. These bands are defined modulus $2\pi$ and symmetric with respect to $E=0$. So they can only meet with each other at $E_\pm=0$ or $E_\pm=\pm\pi$. We thus find the gapless condition of the Floquet spectrum to be $\cos(K_1\cos k)\cos(K_2\sin k)=\pm1$. In the main text, we focus on the case with the kicking strength $K_1=0.5\pi$. In this case, the gapless condition can be satisfied only if $K_1\cos k=0$, i.e., $k=\pm\pi/2$, enforcing $K_2$ to be integer multiples of $\pi$. More specifically, when $K_1=0.5\pi$ and $K_2=(2{\mathbb Z})\pi$ [$K_2=(2{\mathbb Z}-1)\pi$], the two Floquet bands touch at $E_\pm(k=\pm\pi/2)=0$ [$E_\pm(k=\pm\pi/2)=\pm\pi$]. In both cases, the quasienergy gap closes at two points $k=\pm\pi/2$ in $k$-space when a transition happens. The Floquet spectra of the spin-$1/2$ ORDKR for $K_1=0.5\pi$ and $K_2=\pi,2\pi,3\pi,4\pi$ are presented in Fig.~\ref{fig:KRS-Evsk} in order to verify our analysis.

\begin{figure}
	\begin{centering}
		\includegraphics[scale=0.5]{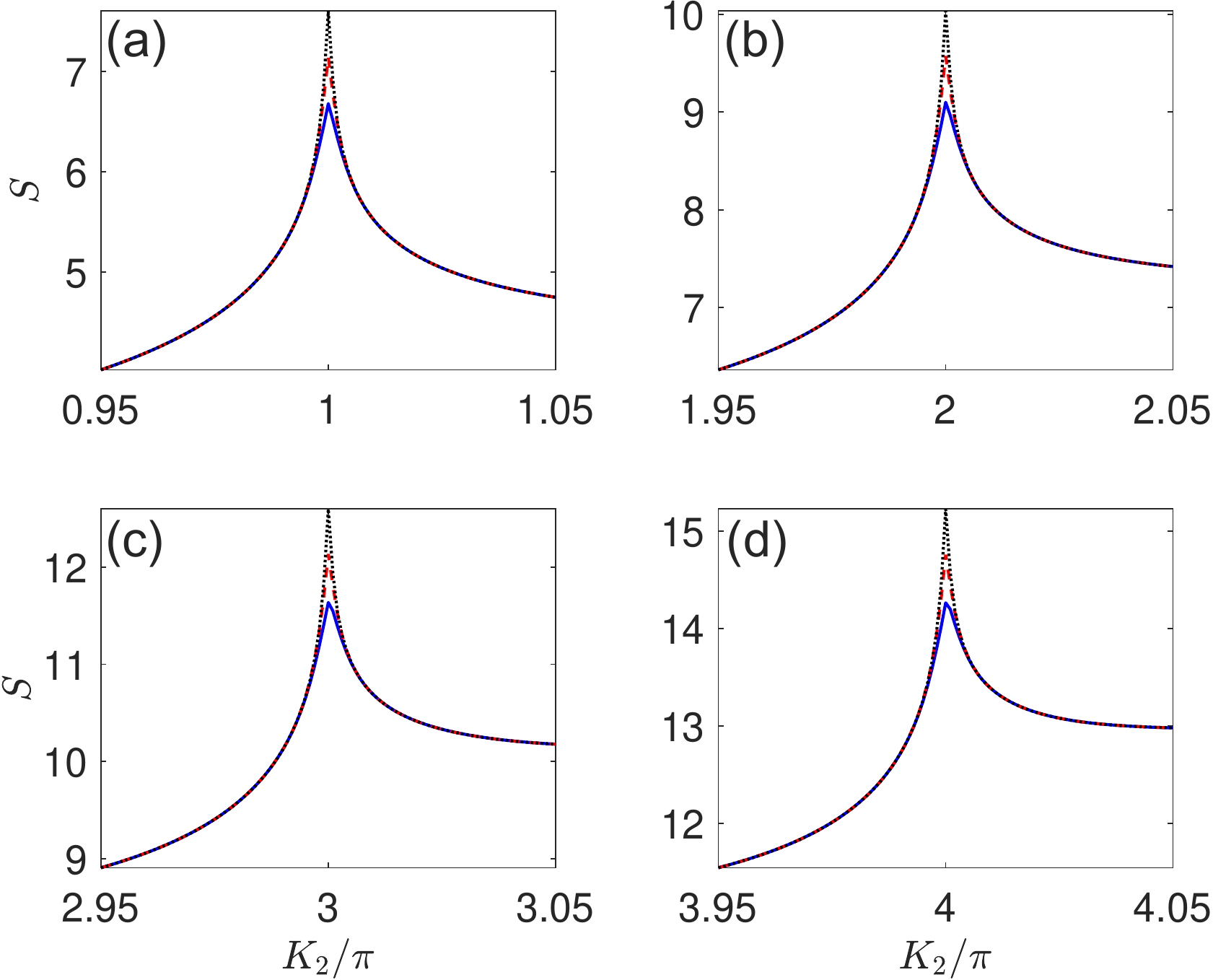}
		\par\end{centering}
	\caption{EE of the spin-$1/2$ ORDKR $S$ versus $K_2$ at different lattice sizes $L$. The size of subsystem A is $L/2$. We set $K_1=0.5\pi$ for all cases. In each panel, the blue solid, red dashed and black dotted lines represent the EE at the lattice sizes $L=1000$, $2000$ and $4000$, respectively.}\label{fig:KRS-EE}
\end{figure}

\begin{figure}
	\begin{centering}
		\includegraphics[scale=0.5]{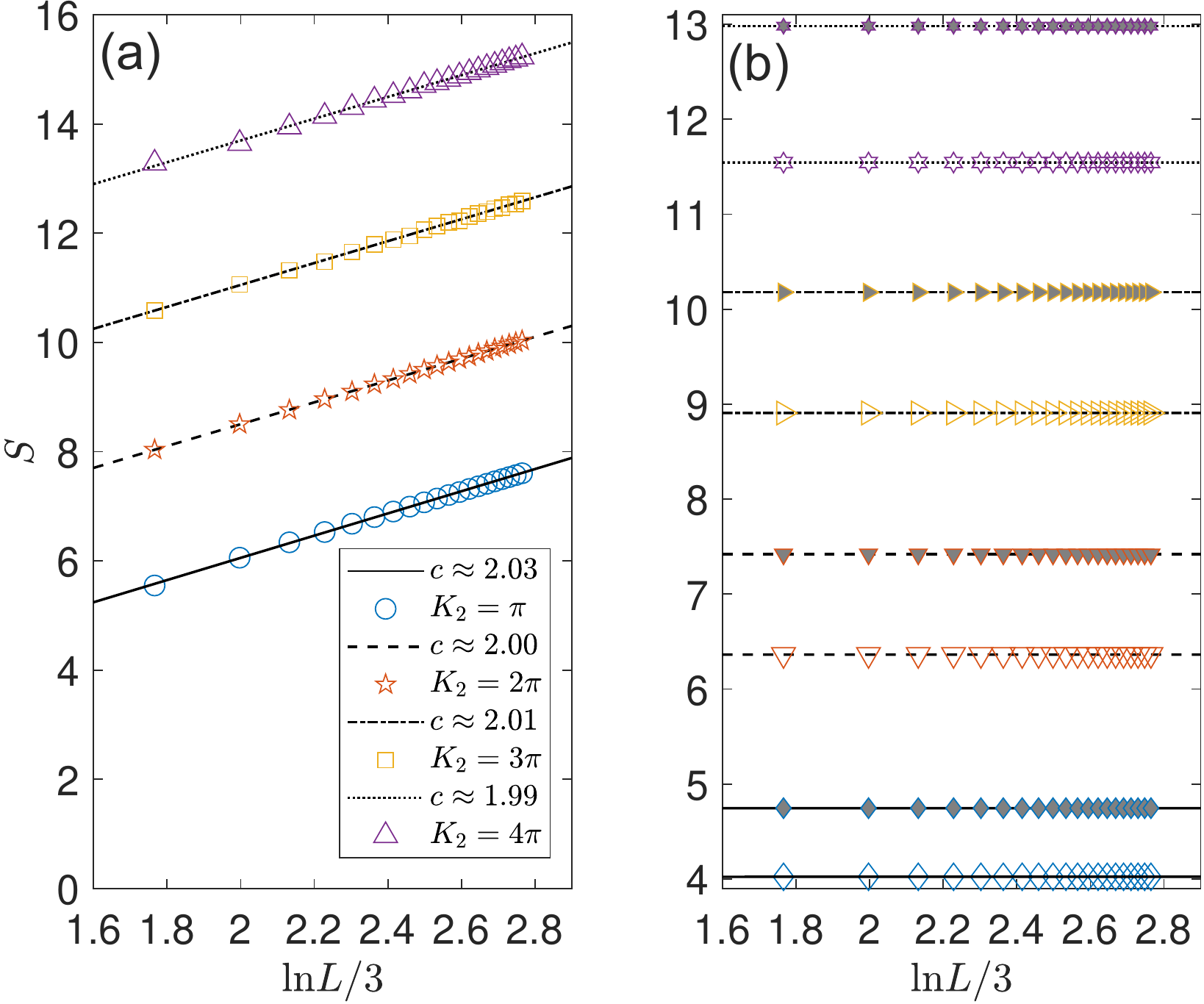}
		\par\end{centering}
	\caption{EE of the spin-$1/2$ ORDKR $S$ versus the lattice size $L$. The size of subsystem A is $L/2$. We set $K_1=0.5\pi$ for all cases and let $L$ goes from $200$ up to $4000$ for each data set. In panel (a), $c$ refers to the central charge extracted from the linear fitting of each data set of $S$ at different $K_2$. In panel (b), the data sets from bottom to top have the values of $K_2=0.95\pi$, $1.05\pi$, $1.95\pi$, $2.05\pi$, $2.95\pi$, $3.05\pi$, $3.95\pi$ and $4.05\pi$.}\label{fig:KRS-EEvsL}
\end{figure}

We now discuss the properties of EE around these topological phase transition points. Let us first consider the change of EE with system size. In Fig.~\ref{fig:KRS-EE}, we show the EE around the four topological transition points in Fig.~\ref{fig:KRS-Evsk}. We observe that away from these transition points, the EE almost has no changes with the increse of lattice size. Meanwhile, right at each transition point, the EE raises with the increase of the length of lattice. These observations suggest that the EE has different scaling behaviors versus the system size away from and right at each topological phase transition point. To further clarify these scaling properties, we report the EE at different lattice sizes together with their fitting curves in Fig.~\ref{fig:KRS-EEvsL}. Away from the transition points, we observe that indeed the EE is almost invariant with the increase of the lattice size $L$. Right at each transition point, the leading term in EE is found to scale linearly with $\ln L$. The gradient has the form of $c/3$, with $c$ usually being called the central charge. For the transition with a gap-closing at either $E=0$ or $E=\pm\pi$, we find $c=2$ from Fig.~\ref{fig:KRS-EEvsL}(a). This is consistent with what one may expect in a usual static system with a gapless spectrum \cite{EEGapless1,EEGapless2,EEGapless3}, since at half-filling, there are only two occupied states at the two band-touching points for all the cases shown in Fig.~\ref{fig:KRS-Evsk}. A similar result for the finite-size scaling of EE was reported before in a 1D Floquet topological superconductor \cite{YatesPRL2018}. The scaling relation of EE observed here for our 1D Floquet topological insulator coincides with that found in the superconducting setup, where the topological edge states exhibit themselves as Majorana modes~\cite{YatesPRL2018} instead of normal fermionic zero and $\pi$ modes in our case.



\begin{thebibliography}{99}
	
	\bibitem{FloRev1} J. Cayssol, B. D\'ora, F. Simon, and R. Moessner,
	Floquet topological insulators, Phys. Status Solidi RRL \textbf{7},
	101-108 (2013).
	
	\bibitem{FloRev2} A. Eckardt, \emph{Colloquium}: Atomic quantum gases
	in periodically driven optical lattices, Rev. Mod. Phys. \textbf{89},
	011004 (2017).
	
	\bibitem{FloRev3} T. Oka and S. Kitamura, Floquet Engineering of
	Quantum Materials, Annu. Rev. Condens. Matter Phys. \textbf{10}, 387-408
	(2019).
	
	\bibitem{FloRev4} F. Harper, R. Roy, M. S. Rudner, and S. L. Sondhi,
	Topology and Broken Symmetry in Floquet Systems, Annu. Rev. Condens.
	Matter Phys. \textbf{11}, 345-368 (2020).
	
	\bibitem{FloRev5} M. S. Rudner and N. H. Lindner, Band structure
	engineering and non-equilibrium dynamics in Floquet topological insulators,
	Nat. Rev. Phys.\textbf{ 2}, 229 (2020).
	
	\bibitem{FloBigTN1} D. Y. H. Ho and J. Gong, Quantized Adiabatic
	Transport In Momentum Space, Phys. Rev. Lett. \textbf{109}, 010601
	(2012).
	
	\bibitem{FloBigTN2} Q. Tong, J. An, J. Gong, H. Luo, and C. H. Oh,
	Generating many Majorana modes via periodic driving: A superconductor
	model, Phys. Rev. B \textbf{87}, 201109(R) (2013).
	
	\bibitem{FloBigTN3} D. Y. H. Ho and J. Gong, Topological effects
	in chiral symmetric driven systems, Phys. Rev. B \textbf{90}, 195419
	(2014).
	
	\bibitem{FloBigTN4} L. Zhou, H. Wang, D. Y. H. Ho, and J. Gong, Aspects
	of Floquet bands and topological phase transitions in a continuously
	driven superlattice, Eur. Phys. J. B \textbf{87}, 204 (2014).
	
	\bibitem{FloBigTN5} M. Lababidi, I. I. Satija, and E. Zhao, Counter-propagating
	Edge Modes and Topological Phases of a Kicked Quantum Hall System,
	Phys. Rev. Lett. \textbf{112}, 026805 (2014).
	
	\bibitem{FloBigTN6} Z. Zhou, I. I. Satija, and E. Zhao, Floquet edge
	states in a harmonically driven integer quantum Hall system, Phys.
	Rev. B \textbf{90}, 205108 (2014).
	
	\bibitem{FloBigTN7} T. Xiong, J. Gong, and J. An, Towards large-Chern-number
	topological phases by periodic quenching, Phys. Rev. B \textbf{93},
	184306 (2016).
	
	\bibitem{FloBigTN8} H. H. Yap, L. Zhou, C. H. Lee, and J. Gong, Photoinduced
	half-integer quantized conductance plateaus in topological-insulator/superconductor
	heterostructures, Phys. Rev. B \textbf{97}, 165142 (2018).
	
	\bibitem{FloBigTN9} R. W. Bomantara, L. Zhou, J. Pan, and J. Gong,
	Coupled-wire construction of static and Floquet second-order topological
	insulators, Phys. Rev. B \textbf{99}, 045441 (2019).
	
	\bibitem{FloBigTN10} L. Zhou, R. W. Bomantara, and S. Wu, $q$th-root non-Hermitian Floquet topological insulators, SciPost Phys. {\bf 13}, 015 (2022).
	
	\bibitem{FloClass1} F. Nathan and M. S. Rudner, Topological singularities
	and the general classification of Floquet-Bloch systems, New J. Phys.\textbf{17},
	125014 (2015).
	
	\bibitem{FloClass2} A. C. Potter, T. Morimoto, and A. Vishwanath,
	Classification of Interacting Topological Floquet Phases in One Dimension,
	Phys. Rev. X \textbf{6}, 041001 (2016). 
	
	\bibitem{FloClass3} R. Roy and F. Harper, Periodic table for Floquet
	topological insulators, Phys. Rev. B \textbf{96}, 155118 (2017).
	
	\bibitem{FloClass4} S. Yao, Z. Yan, and Z. Wang, Topological invariants
	of Floquet systems: General formulation, special properties, and Floquet
	topological defects, Phys. Rev. B \textbf{96}, 195303 (2017).
	
	\bibitem{FloAnoES1} M. S. Rudner, N. H. Lindner, E. Berg, and M.
	Levin, Anomalous Edge States and the Bulk-Edge Correspondence for
	Periodically Driven Two-Dimensional Systems, Phys. Rev. X \textbf{3},
	031005 (2013).
	
	\bibitem{FloAnoES2} P. Titum, E. Berg, M. S. Rudner, G. Refael, and
	N. H. Lindner, Anomalous Floquet-Anderson Insulator as a Nonadiabatic
	Quantized Charge Pump, Phys. Rev. X \textbf{6}, 021013 (2016).
	
	\bibitem{FloAnoES3} L. Zhou and J. Gong, Recipe for creating an arbitrary
	number of Floquet chiral edge states, Phys. Rev. B \textbf{97}, 245430
	(2018).
	
	\bibitem{FloExp1} T. Kitagawa, M. A. Broome, A. Fedrizzi, M. S. Rudner,
	E. Berg, I. Kassal, A. Aspuru-Guzik, E. Demler, and A. G. White, Observation
	of topologically protected bound states in photonic quantum walks,
	Nat. Commun.\textbf{3}, 882 (2012).
	
	\bibitem{FloExp2} M. C. Rechtsman, J. M. Zeuner, Y. Plotnik, Y. Lumer,
	D. Podolsky, F. Dreisow, S. Nolte, M. Segev, and A. Szameit, Photonic
	Floquet topological insulators, Nature \textbf{496}, 196-200 (2013).
	
	\bibitem{FloExp3} Y. Wang, H. Steinberg, P. Jarillo-Herrero, and
	N. Gedik, Observation of Floquet-Bloch States on the Surface of a
	Topological Insulator, Science \textbf{342}, 453-457 (2013).
	
	\bibitem{FloExp4} G. Jotzu, M. Messer, R. Desbuquois, M. Lebrat,
	T. Uehlinger, D. Greif, and T. Esslinger, Experimental realization
	of the topological Haldane model with ultracold fermions, Nature \textbf{515},
	237-240 (2014).
	
	\bibitem{FloExp5} W. Hu, J. C. Pillay, K. Wu, M. Pasek, P. P. Shum,
	and Y. D. Chong, Measurement of a Topological Edge Invariant in a
	Microwave Network, Phys. Rev. X \textbf{5}, 011012 (2015).
	
	\bibitem{FloExp6} L. Asteria, D. T. Tran, T. Ozawa, M. Tarnowski,
	B. S. Rem, N. Fl\"aschner, K. Sengstock, N. Goldman, and C. Weitenberg,
	Measuring quantized circular dichroism in ultracold topological matter,
	Nat. Phys. \textbf{15}, 449-454 (2019).
	
	\bibitem{FloExp7} K. Wintersperger, C. Braun, F. N. \"Unal, A. Eckardt,
	M. D. Liberto, N. Goldman, I. Bloch, and M. Aidelsburger, Realization
	of an anomalous Floquet topological system with ultracold atoms, Nat.
	Phys.\textbf{ 16}, 1058-1063 (2020).
	
	\bibitem{FloExp8} J. W. McIver, B. Schulte, F.-U. Stein, T. Matsuyama,
	G. Jotzu, G. Meier, and A. Cavalleri, Light-induced anomalous Hall
	effect in graphene, Nat. Phys. \textbf{16}, 38-41 (2020).
	
	\bibitem{FloQC1} R. W. Bomantara and J. Gong, Simulation of Non-Abelian
	Braiding in Majorana Time Crystals, Phys. Rev. Lett. \textbf{120},
	230405 (2018).
	
	\bibitem{FloQC2} R. W. Bomantara and J. Gong, Quantum computation
	via Floquet topological edge modes, Phys. Rev. B \textbf{98}, 165421
	(2018).
	
	\bibitem{FloQC3} R. W. Bomantara and J. Gong, Measurement-only quantum
	computation with Floquet Majorana corner modes, Phys. Rev. B \textbf{101},
	085401 (2020).
	
	\bibitem{EntangleRev1} L. Amico, R. Fazio, A. Osterloh, and V. Vedral,
	Entanglement in many-body systems, Rev. Mod. Phys. \textbf{80}, 517
	(2008).
	
	\bibitem{EntangleRev2} R. Horodecki, P. Horodecki, M. Horodecki,
	and K. Horodecki, Quantum entanglement, Rev. Mod. Phys. \textbf{81},
	865 (2009).
	
	\bibitem{EntangleRev3} J. I. Latorre and A. Riera, A short review
	on entanglement in quantum spin systems, J. Phys. A: Math. Theor.
	\textbf{42}, 504002 (2009).
	
	\bibitem{EntangleRev4} I. Peschel and V. Eisler, Reduced density
	matrices and entanglement entropy in free lattice models, J. Phys.
	A: Math. Theor. \textbf{42}, 504003 (2009).
	
	\bibitem{EntangleRev5} J. Eisert, M. Cramer, and M. B. Plenio, \emph{Colloquium}:
	Area laws for the entanglement entropy, Rev. Mod. Phys. \textbf{82},
	277 (2010).
	
	\bibitem{EntangleRev6} T. Nishioka, Entanglement entropy: Holography and renormalization group, Rev. Mod. Phys. {\bf 90}, 035007 (2018). 
	
	\bibitem{EntangleRev7} M. Erhard, M. Krenn, and A. Zeilinger, Advances
	in high-dimensional quantum entanglement, Nat. Rev. Phys. \textbf{2},
	365-381 (2020).
	
	\bibitem{ES0} H. Li and F. D. M. Haldane, Entanglement Spectrum as
	a Generalization of Entanglement Entropy: Identification of Topological
	Order in Non-Abelian Fractional Quantum Hall Effect States, Phys.
	Rev. Lett. \textbf{101}, 010504 (2008).
	
	\bibitem{EE0} M. Haque, O. Zozulya, and K. Schoutens, Entanglement
	Entropy in Fermionic Laughlin States, Phys. Rev. Lett. \textbf{98},
	060401 (2007).
	
	\bibitem{ESEETP1} G. Vidal, J. I. Latorre, E. Rico, and A. Kitaev,
	Entanglement in Quantum Critical Phenomena, Phys. Rev. Lett. \textbf{90},
	227902 (2003).
	
	\bibitem{ESEETP2} V. E. Korepin, Universality of Entropy Scaling
	in One Dimensional Gapless Models, Phys. Rev. Lett. \textbf{92}, 096402
	(2004).
	
	\bibitem{ESEETP3} M. Levin and X. Wen, Detecting Topological Order
	in a Ground State Wave Function, Phys. Rev. Lett. \textbf{96}, 110405
	(2006).
	
	\bibitem{ESEETP4} A. Kitaev and J. Preskill, Topological Entanglement
	Entropy, Phys. Rev. Lett. \textbf{96}, 110404 (2006).
	
	\bibitem{ESEETP42} Y. Zhang, T. Grover, and A. Vishwanath, Topological
	entanglement entropy of $\mathbb{Z}_{2}$ spin liquids and lattice
	Laughlin states, Phys. Rev. B \textbf{84}, 075128 (2011).
	
	\bibitem{ESEETP5} L. Fidkowski, Entanglement Spectrum of Topological
	Insulators and Superconductors, Phys. Rev. Lett. \textbf{104}, 130502
	(2010).
	
	\bibitem{ESEETP52} H. Yao and X.-L. Qi, Entanglement Entropy and Entanglement Spectrum of the Kitaev Model,
	Phys. Rev. Lett. {\bf 105}, 080501 (2010).
	
	\bibitem{ESEETP6} E. Prodan, T. L. Hughes, and B. A. Bernevig, Entanglement
	Spectrum of a Disordered Topological Chern Insulator, Phys. Rev. Lett.
	\textbf{105}, 115501 (2010).
	
	\bibitem{ESEETP7} A. Alexandradinata, T. L. Hughes, and B. A. Bernevig,
	Trace index and spectral flow in the entanglement spectrum of topological
	insulators, Phys. Rev. B \textbf{84}, 195103 (2011).
	
	\bibitem{ESEETP8} X. Qi, H. Katsura, and A. W. W. Ludwig, General
	Relationship between the Entanglement Spectrum and the Edge State
	Spectrum of Topological Quantum States, Phys. Rev. Lett. \textbf{108},
	196402 (2012).
	
	\bibitem{ESEETP9} Z. Huang and D. P. Arovas, Entanglement spectrum
	and Wannier center flow of the Hofstadter problem, Phys. Rev. B \textbf{86},
	245109 (2012).
	
	\bibitem{ESEETP10} M. Legner and T. Neupert, Relating the entanglement
	spectrum of noninteracting band insulators to their quantum geometry
	and topology, Phys. Rev. B \textbf{88}, 115114 (2013).
	
	\bibitem{ESEETP11} T. H. Hsieh and L. Fu, Bulk Entanglement Spectrum
	Reveals Quantum Criticality within a Topological State, Phys. Rev.
	Lett. \textbf{113}, 106801 (2014).
	
	\bibitem{ESEETP12} M. Hermanns, Y. Salimi, M. Haque and L. Fritz,
	Entanglement spectrum and entanglement Hamiltonian of a Chern insulator
	with open boundaries, J. Stat. Mech. \textbf{2014}, P10030 (2014).
	
	\bibitem{ESEETP13} G. B. Hal\'asz and A. Hamma, Topological R\'enyi Entropy
	after a Quantum Quench, Phys. Rev. Lett. \textbf{110}, 170605 (2013).
	
	\bibitem{YatesPRB2016} D. J. Yates, Y. Lemonik, and A. Mitra, Entanglement
	properties of Floquet-Chern insulators, Phys. Rev. B \textbf{94},
	205422 (2016).
	
	\bibitem{YatesPRB2017} D. J. Yates and A. Mitra, Entanglement properties
	of the time-periodic Kitaev chain, Phys. Rev. B \textbf{96}, 115108
	(2017).
	
	\bibitem{YatesPRL2018} D. J. Yates, Y. Lemonik, and A. Mitra, Central
	Charge of Periodically Driven Critical Kitaev Chains, Phys. Rev. Lett.
	\textbf{121}, 076802 (2018).
	
	\bibitem{JafariPRA2021} R. Jafari and A. Akbari, Floquet dynamical phase transition and entanglement spectrum, Phys. Rev. A {\bf 103}, 012204 (2021).
	
	\bibitem{OBWN1} A. Kitaev, Anyons in an exactly solved model and
	beyond, Ann. Phys. (Amsterdam) \textbf{321}, 2 (2006).
	
	\bibitem{OBWN2} E. Prodan, Non-commutative tools for topological
	insulators, New J. Phys. \textbf{12}, 065003 (2010).
	
	\bibitem{OBWN3} I. Mondragon-Shem, T. L. Hughes, J. Song, and E. Prodan,
	Topological Criticality in the Chiral-Symmetric AIII Class at Strong Disorder,
	Phys. Rev. Lett. {\bf 113}, 046802 (2014).
	
	\bibitem{OBWN4} L. Zhou, Y. Gu, and J. Gong, Dual topological characterization
	of non-Hermitian Floquet phases, Phys. Rev. B \textbf{103}, L041404
	(2021).
	
	\bibitem{RSCN1} R. Bianco and R. Resta, Mapping topological order
	in coordinate space, Phys. Rev. B \textbf{84}, 241106(R) (2011).
	
	\bibitem{RSCN2} A. Marrazzo and R. Resta, Locality of the anomalous
	Hall conductivity, Phys. Rev. B \textbf{95}, 121114(R) (2017).
	
	\bibitem{RSCN3} D. T. Tran, A. Dauphin, A. G. Grushin, P. Zoller, and N. Goldman, Probing topology by ``heating'': Quantized circular dichroism in ultracold atoms, Sci. Adv. {\bf 3}, e1701207 (2017).
	
	\bibitem{RSCN4} M. Davide Caio, G. M\"oller, N. R. Cooper, and M. J. Bhaseen, Topological marker currents in Chern insulators, Nat. Phys. \textbf{15}, 257 (2019).
	
	\bibitem{RSCN5} B. Irsigler, J.-H. Zheng, and W. Hofstetter, Microscopic characteristics and tomography scheme of the local Chern marker, Phys. Rev. A {\bf 100}, 023610 (2019).
	
	\bibitem{KRSZhou2018} L. Zhou and J. Gong, Floquet topological phases
	in a spin-$1/2$ double kicked rotor, Phys. Rev. A \textbf{97}, 063603
	(2018).
	
	\bibitem{KRSChen2021} B. Chen, S. Li, X. Hou, F. Ge, F. Zhou, P.
	Qian, F. Mei, S. Jia, N. Xu, and H. Shen, Digital quantum simulation
	of Floquet topological phases with a solid-state quantum simulator,
	Photon. Res. \textbf{9}, 81-87 (2021).
	
	\bibitem{KRSBolik2022} N. Bolik, C. Groiseau, J. H. Clark, G. S. Summy, Y. Liu, and S. Wimberger,
	Detecting topological phase transitions in a double kicked quantum rotor, Phys. Rev. A {\bf 106}, 043318 (2022).
	
	\bibitem{AsbothSTM} J. K. Asb\'oth and H. Obuse, Bulk-boundary correspondence
	for chiral symmetric quantum walks, Phys. Rev. B \textbf{88}, 121406(R)
	(2013).
	
	\bibitem{SCLZhou2020} L. Zhou and Q. Du, Floquet topological phases
	with fourfold-degenerate edge modes in a driven spin-$1/2$ Creutz
	ladder, Phys. Rev. A \textbf{101}, 033607 (2020).
	
	\bibitem{WangPRE2013} H. Wang, D. Y. H. Ho, W. Lawton, J. Wang, and
	J. Gong, Kicked-Harper model versus on-resonance double-kicked rotor
	model: From spectral difference to topological equivalence, Phys.
	Rev. E \textbf{88}, 052920 (2013).
	
	\bibitem{Haldane1} F. D. M. Haldane, Model for a Quantum Hall Effect
	without Landau Levels: Condensed-Matter Realization of the ``Parity
	Anomaly'', Phys. Rev. Lett. \textbf{61}, 2015 (1988).
	
	\bibitem{Haldane2} C. Bena and L. Simon, Dirac point metamorphosis
	from third-neighbor couplings in graphene and related materials, Phys.
	Rev. B \textbf{83}, 115404 (2011).
	
	\bibitem{Haldane3} D. Sticlet and F. Piéchon, Chern mosaic: Topology
	of chiral superconductivity on ferromagnetic adatom lattices, Phys.
	Rev. B \textbf{87}, 115402 (2013).
	
	\bibitem{LargeCNExp} K. Yang, S. Xu, L. Zhou, Z. Zhao, T. Xie, Z.
	Ding, W. Ma, J. Gong, F. Shi, and J. Du, Observation of Floquet topological phases with large Chern numbers, arXiv:2209.05275.
	
	\bibitem{EETPT1} M. Rodney, H. F. Song, S.-S. Lee, K. L. Hur, and E. S. Sørensen, Scaling of entanglement entropy across Lifshitz transitions,
	Phys. Rev. B {\bf 87}, 115132 (2013).
	
	\bibitem{EETPT2} T. P. Oliveira, P. Ribeiro, and P. D. Sacramento, Entanglement entropy and entanglement spectrum of triplet topological superconductors,
	J. Phys.: Condens. Matter {\bf 26}, 425702 (2014).
	
	\bibitem{EETPT3} J. Cho and K. W. Kim, Quantum Phase Transition and Entanglement in Topological Quantum Wires, Sci. Rep. {\bf 7}, 2745 (2017).
	
	\bibitem{EETPT4} R. Nehra, D. S. Bhakuni, S. Gangadharaiah, and A. Sharma, Many-body entanglement in a topological chiral ladder, Phys. Rev. B {\bf 98}, 045120 (2018).
	
	\bibitem{WSM1} R. W. Bomantara, G. N. Raghava, L. Zhou, and J. Gong, Floquet topological semimetal phases of an extended kicked Harper model, Phys. Rev. E {\bf 93}, 022209 (2016).
	
	\bibitem{WSM2} H. Wang, L. Zhou, and Y. D. Chong, Floquet Weyl phases in a three-dimensional network model, Phys. Rev. B {\bf 93}, 144114 (2016).
	
	\bibitem{KRSOther1} J. P. Dahlhaus, J. M. Edge, J. Tworzydlo, and C. W. J. Beenakker, Quantum Hall effect in a one-dimensional dynamical system,
	Phys. Rev. B {\bf 84}, 115133 (2011).
	
	\bibitem{KRSOther2} I. Dana and K. Kubo, Floquet systems with Hall effect: Topological properties and phase transitions, Phys. Rev. B {\bf 100}, 045107 (2019).
	
	\bibitem{KRSTian1} Y. Chen and C. Tian, Planck's Quantum-Driven Integer Quantum Hall Effect in Chaos, Phys. Rev. Lett. {\bf 113}, 216802 (2014).
	
	\bibitem{KRSTian2} C. Tian, Y. Chen, and J. Wang, Emergence of integer quantum Hall effect from chaos, Phys. Rev. B {\bf 93}, 075403 (2016).
	
	\bibitem{INTQKR1} S. Lellouch, A. Rançon, S. De Bi\`evre, D. Delande, and J. C. Garreau, Dynamics of the mean-field-interacting quantum kicked rotor, Phys. Rev. A {\bf 101}, 043624 (2020).
	
	\bibitem{INTQKR2} R. Chicireanu and A. Rançon, Dynamical localization of interacting bosons in the few-body limit, Phys. Rev. A {\bf 103}, 043314 (2021).
	
	\bibitem{EEGapless1} B.-Q. Jin and V. E. Korepin, Quantum Spin Chain, Toeplitz Determinants and the Fisher-Hartwig Conjecture, Journal of Statistical Physics {\bf 116}, 79-95 (2004).
	
	\bibitem{EEGapless2} P. Calabrese and J. Cardy, Entanglement entropy and quantum field theory, J. Stat. Mech. {\bf 2004}, P06002 (2004).
	
	\bibitem{EEGapless3} B. Swingle, Entanglement Entropy and the Fermi Surface, Phys. Rev. Lett. {\bf 105}, 050502 (2010).
	
	
	
\end{thebibliography}
\end{document}